\documentclass[aps,prb,twocolumn,groupedaddress,floatfix,superscriptaddress]{revtex4-1}
\usepackage{epsfig}
\usepackage{amsmath,amssymb,graphicx,ulem}
\usepackage{physics}
\usepackage[dvipsnames,usenames]{xcolor}

\begin{document}

\title{Designer Flat Bands: Topology and Enhancement of Superconductivity} 

\author{Si Min Chan}

\affiliation{Centre for Quantum Technologies, National University of
  Singapore, 3 Science Drive 2, 117543 Singapore}

\affiliation{Department of Physics, National University of Singapore,
  2 Science Drive 3, 117551 Singapore}

\author{B. Gr\'emaud}

\affiliation{Aix Marseille Univ, Université de Toulon, CNRS, CPT,
  IPhU, AMUTech, Marseille, France}
 
\author{G. G. Batrouni}

\affiliation{Centre for Quantum Technologies, National University of
  Singapore, 3 Science Drive 2, 117543 Singapore}

\affiliation{Department of Physics, National University of Singapore,
  2 Science Drive 3, 117551 Singapore}

\affiliation{Beijing Computational Science Research Center, Beijing
  100193, China}

\affiliation{Universit\'e C\^ote d'Azur, CNRS, Institut de Physique de
  Nice (INPHYNI), 06103 Nice, France}

\begin{abstract}

We construct quasi one-dimensional topological and non-topological
three-band lattices with tunable band gap and winding number of the
flat band. Using mean field (MF) and exact density matrix
renormalization group (DMRG) calculations, we show explicitly how the
band gap affects pairing and superconductivity (SC) in a Hubbard model
with attractive interactions. We show excellent agreement between MF
and DMRG. When a phase twist is applied on the system, a phase
difference appears between pairing order parameters on different
sublattices, and this plays a very important role in the SC
density. The SC weight, $D_s$, on the gapped topological, $W\neq0$,
flat band increases linearly with interaction strength, $U$, for low
values, and with a slope that depends on the details of the compact
localized state at $U=0$. As $U\to 0$ for the gapped non-topological
flat band ($W=0$), $D_s$ decays with a power law faster than quadratic
but slower than exponential. This indicates that isolated
non-topological flat bands are less beneficial to SC.  In the gapless
case (flat band touching the band above it), we find at low $U$ (both
for $W=0$ and $W\neq 0$) that $D_s\propto U^\varphi$ with $\varphi<1$
contrary to the $U{\rm ln}\, ({\rm const.}/U)$ behavior reported in
the literature. In other words, $D_s$ increases faster than linearly
for low $U$ thus favoring SC at weak interaction more than the gapped
case. For systems with touching bands, we observe that the one-body
correlation length, $\xi$, diverges as a power law as $U\rightarrow0$,
while for the isolated flat band $\xi(U\to 0)$ is a constant smaller
than one lattice spacing. Both behaviors are distinct from the
exponential divergence of $\xi$ in the dispersive case. Our results
re-establish that the BCS mean field and quantum metric alone are
insufficient to characterize SC at weak coupling.

\end{abstract}

\maketitle 

\section{Introduction}

Flat band physics has garnered wide interest since the 1990s, but this
captivation has become exceptionally pronounced following the
experimental realization of unconventional superconductivity in
twisted bilayer graphene at a ``magic
angle''\cite{cao2018unconventional,cao18,yankowitz19,park21} where the
appearance of a flat band is suspected to be the driving
mechanism. The possibility of revealing exotic quantum phases --- in
particular, superconductivity on topological and non-topological flat
bands --- can be attributed to a unique characteristic of these
dispersionless bands, where any finite interaction will be much larger
than the band width, leading to strongly correlated physics at any
value of the interaction.

When a particle is loaded in a flat band, the high degeneracy causes
it to localize in a compact form within a few sites whose geometry
depends on the details of the Hamiltonian; we will refer to this as
the compact localized state (CLS).  Studies on topological models have
argued that, at weak coupling, isolated flat bands enhance pair
formation and superconductivity (SC) and raise the BCS transition
temperature,
$T_c$. \cite{khodel1990superfluidity,peotta2015superfluidity,kopnin2011high}
It was demonstrated, with computational and mean field methods, that a
partially filled isolated flat band has superfluid weight, $D_s$,
linear in the interaction, $U$, for $U$ much smaller than the gap,
where transport is dominated by the topology of the flat
band. \cite{peotta2015superfluidity,mondaini2018pairing,chan2022pairing,huhtinen2022revisiting,tovmasyan2018preformed}
Furthermore, the slope at linearity is not simply given by the quantum
metric but is accounted for, very accurately, by a proper projection
on the flat band taking into account the inequivalent
sublattices.\cite{chan2022pairing} Recently, the quantum metric
prediction for the slope was improved by introducing the notion of
minimal quantum metric.\cite{huhtinen2022revisiting} For the sawtooth
lattice at a filling of $\rho = 0.5$, the quantum metric predicts the
slope of $D_s$ to be $0.6$ compared to exact DMRG, multi-band MF, and
proper projection on the flat band, all of which yield slopes of
$0.40$.\cite{chan2022pairing} The minimal quantum
metric\cite{huhtinen2022revisiting} gives a slope of $0.45$ which
brings it closer to the exact DMRG and MF found in
Ref.[\onlinecite{chan2022pairing}]. In dimensions two and above, this
linear behavior of $D_s$ with $U$ was shown to lead to similarly
linear dependence of $T_c$ on the
coupling.\cite{pyykkonen21,heikkila2011flat,kopnin2011high,khodel1990superfluidity,miyahara2007bcs}
We recall that in dispersive bands, $D_s$ and $T_c$ are exponentially
small as $U\rightarrow 0$, with $D_s \sim {\rm e}^{-a/U}$ and $T_c\sim
{\rm e}^{-b/U}$.

Full multi-band mean field (MF) methods can accurately describe the
entire range of superconducting behavior, from weak to very strong
interactions where particles behave effectively like hard-core bosons
on a dispersive band. Despite the changing transport mechanisms as $U$
is increased, the full multi-band MF method, which accounts for
sublattice inequivalence, faithfully recovers correct results across
the entire range of $U$.\cite{chan2022pairing}

When the lowest energy flat band just touches the next dispersive
band, any finite interaction will necessarily involve both bands in
transport. It has been suggested, using the BCS mean field, quantum
metric, and the minimal quantum metric, that touching bands can be
beneficial to superconductivity
\cite{huhtinen2022revisiting,iskin2019origin,wu2021superfluid,julku2016geometric}
resulting in $D_s\propto U{\rm ln}\, ({\rm const.}/U)$.

In this work, we address two main questions. First, what is the role
played in SC by topology as opposed to band flatness? In other words,
suppose we have two systems with very similar-looking band structures:
the lowest band is flat and separated from the next band above it, but
in one system the flat band has nonzero winding number and in the
other $W=0$. We examine how superconductivity differs in these two
systems, and further consider its dependence on the CLS of the flat
band.

The second question is: How do the answers to the first question
change when the flat band touches the band above it, i.e. in the
gapless case? Both these questions are addressed in the case where the
flat band filling is less than full.

Such questions have been addressed previously. However, in the case of
comparing topological and non-topological flat bands, the systems that
were compared had different structures. For example, in
Ref.[\onlinecite{verma21}], the non-topological Lieb lattice was
compared with the topological $\pi$-flux lattice; the former is a
three-band system with the flat band in between two dispersive bands,
whereas the latter is a two-band system with the flat band in the
ground state. So, the systems are quite different; it would be
instructive to compare two very similar systems but with different
topological properties.

To this end, we first focus on a quasi one-dimensional three-band
(i.e. three-orbital) system, where the winding number, $W$, the
filling on the CLS, the flat band energy, and the band gap can all be
tuned by engineering the hopping parameters. We accomplish this by
applying the method of Ref.[\onlinecite{maimaiti2019universal}]. We
show how to obtain the Hubbard Hamiltonian for a general three-band
system, and the full multi-band mean field required to describe these
systems accurately. We outline the construction of new systems with
the desired winding number, and with the flat band as the lowest
energy state, and the CLS on two neighboring unit cells. With the
chosen $W$, CLS, and flat band energy, we still have the additional
freedom to tune the gap through free parameters controlling the next
dispersive band. We note that flat band systems have been realized
experimentally with photonic lattices in two
dimensions\cite{xia16,zong16} and in quasi
one-dimension.\cite{mukherjee15,mukherjee17,baboux16}

With these tools in hand, our main results are as follows. For the
gapped topological ($W\neq 0$) system we show that, as in
Ref.[\onlinecite{chan2022pairing}], the full multi-band MF method
agrees very well with exact DMRG in accounting for the properties of
the system over a wide range of coupling parameter and densities. In
particular, we again find that for $U$ smaller than the band gap,
$D_s$ increases linearly with $U$.  At fixed $W$, band gap, and flat
band energy, the relative populations of the two unit cells on which
resides the CLS can be tuned and the largest slope (fastest increase)
of $D_s$ as a function of $U$ is achieved with a symmetric population
on the CLS. We also found that the slope of $D_s(U)$ depends much more
sensitively on the relative populations of the two unit cells than on
the quantum metric. Furthermore, unlike the Creutz and Sawtooth
lattices,\cite{chan2022pairing} we establish, in general, the
dependence of the phase of the order parameters on the interaction
strength, $U$, and band gap. These properties are qualitatively the
same for any $W\neq 0$. For the non-topological $W=0$, we show that a
CLS on two neighboring unit cells cannot be constructed---the CLS now
resides in only one cell. This then implies that transport requires
the upper band. This is confirmed by DMRG and MF calculations which
show that for low $U$, $D_s$ is suppressed and increases slower than
linearly.

When the gap is closed and the flat band touches the band above it,
the upper band will be involved in pairing and SC for any nonzero
$U$. In this case we find that for low $U$ and for both topological
($W=0$) and non-topological ($W\neq 0$) bands, $D_s\propto U^\varphi$
with $\varphi<1$. This means that, for small values of $U$, $D_s$
increases faster than linearly. This power law behavior is in
disagreement with the $U{\rm ln}\, ({\rm const.}/U)$ behavior reported
in the
literature.\cite{huhtinen2022revisiting,iskin2019origin,wu2021superfluid,julku2016geometric}
In this paper we only consider particle densities below full filling
of the flat band. So, in a $3$-band system, we only consider total
densities below $2/3$.

The paper is organized as follows. In section \ref{Model and methods}
we discuss the model Hamiltonian and how we construct it with the
desired CLS and values of $W$, flat band energy and gap. We also
summarize our full multi-band mean field method. In section
\ref{results:nonzero_winding} we discuss the properties of the gapped
and gapless topological systems, $W\neq 0$, while in section
\ref{results:zerowinding} we examine the non-topological, $W=0$,
case. Our conclusions are discussed in section \ref{concs}. Additional
details and results are discussed in five appendices.

\section{Model and methods}\label{Model and methods}

\subsection{Hubbard Hamiltonian: Methods}

\begin{figure}[ht]
  \includegraphics[width=8.6cm]{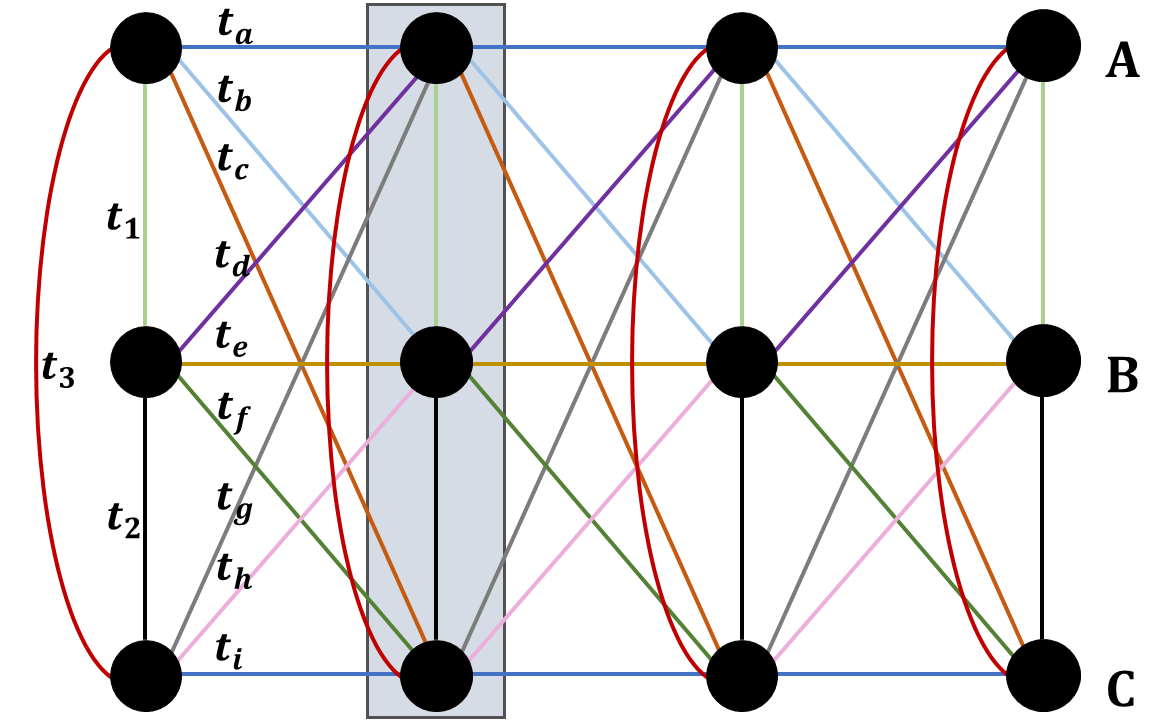}
 \caption{(Color online) The lattice with intra-cell hopping
   parameters $t_1$, $t_2$, $t_3$ and inter-cell hopping parameters
   labeled from $t_a$ to $t_i$. We consider three sublattice sites,
   $A, B, C$, per unit cell (shown in the shaded rectangle) on a quasi
   one-dimensional chain. }
 \label{fig:lattice}
\end{figure}

The Hubbard Hamiltonian with attractive on-site interaction on a
general quasi one-dimensional three-band system is described by
\begin{equation}
\begin{aligned}
    H &= \displaystyle\sum_{i,j,\alpha,\sigma}\left(
    t_{ij}^{\alpha,\alpha'}c^{\alpha\dagger}_{i,\sigma}c^{\alpha'}_{j,\sigma}
    + h.c.\right) -U \displaystyle\sum_{j,\alpha}
    c^{\alpha\dagger}_{j,\downarrow}
    c^{\alpha\dagger}_{j,\uparrow}c^{\alpha}_{j,\uparrow}c^{\alpha}_{j,\downarrow}
    \\
    &-\mu\displaystyle \sum_{j,\alpha,\sigma}
    c^{\alpha\dagger}_{j,\sigma} c^{\alpha}_{j,\sigma},
    \label{eq:hamilt}
\end{aligned}
\end{equation}
where $i,j$ are unit cell labels, $\alpha, \alpha' = A,B,C$ are
sublattice (orbital) indices and $t_{ij}^{\alpha,\alpha'}$ is the
hopping parameter between lattice sites ($i,\alpha$) and ($j,
\alpha'$), as shown in Fig. \ref{fig:lattice}. The operator
$c^{\alpha}_{j,\sigma}$ ($c^{\alpha\dagger}_{j,\sigma}$) destroys
(creates) a spin $\sigma=\uparrow,\downarrow$ fermion on site $i,
\alpha$. The onsite Hubbard interaction parameter between a
spin-$\uparrow$ and spin-$\downarrow$ fermions is $U>0$; $\mu$ is the
chemical potential. Choosing the hopping parameters judiciously (see
below) yields a highly degenerate flat band. We define the filling,
$\rho$, to be the average number of fermions per site, i.e. $\rho =
\frac{N_\uparrow+N_\downarrow}{3L}$ where $L$ is the number of unit
cells.
The superconducting behavior of these systems is probed by applying a
phase gradient $c_{j,\sigma}^\alpha \rightarrow c_{j,\sigma}^\alpha
e^{i\phi j}$, where $\phi$ is the phase gradient $\Phi/L$, and $\Phi$
the phase twist. A defining quantity of SC, the superfluid density
$D_s$, can then be computed from the second derivative of the
thermodynamic grand potential, or equivalently the ground state energy
at zero
temperature\cite{fisher73,zotos90,shastry90,scalapino93,hayward95},
\begin{equation}
    D_s = \pi L \left.\frac{d^2 E_{GS}(\Phi)}{d \Phi^2}\right\vert_{\Phi=0}.
\end{equation}
Superconducting transport in one-dimensional attractive Hubbard
systems has also been shown to be purely pair transport
\cite{gremaud2021pairing} and, consequently, the single particle
Green's function decays exponentially while the pair Green's function
decays as a power:
\begin{align}
\label{eq:GF}
    G^{\alpha\alpha'}_\sigma(r) &= \langle c_{j+r,\sigma}^\alpha
    c_{j,\sigma}^{\alpha'\dagger} \rangle \sim e^{-r/\xi},
    \\
    G^{\alpha\alpha'}_{pair}(r) &= \langle
    c_{j+r,\downarrow}^\alpha c_{j+r,\uparrow}^\alpha
    c_{j,\uparrow}^{\alpha'\dagger}c_{j,\downarrow}^{\alpha'\dagger}
    \rangle \sim r^{-\omega},
\end{align}
where $\xi$ is the correlation length and $\omega$ is the power law
decay exponent.

To study the ground state properties of this system, we use exact DMRG
computation as implemented in the ALPS
library,\cite{albuquerque2007alps, bauer2011alps} and full multi-band
MF. To calculate the Green's functions with DMRG, we use open boundary
conditions (OBC) up to a system size of $L=100$ unit cells. To
calculate $D_s$, we need to apply a phase gradient to induce superflow
which necessitates periodic boundary conditions (PBC). To this end, we
employ the method of Ref.[\onlinecite{mondaini2018pairing}] where
special boundary terms are used which are effectively equivalent to
PBC. In the MF description of the Hubbard Hamiltonian, we decompose
the quartic operator term, $c^{\alpha\dagger}_{j,\downarrow}
c^{\alpha\dagger}_{j,\uparrow} c^{\alpha}_{j,\uparrow}
c^{\alpha}_{j,\downarrow}$, with the full multi-band mean field
expression (Appendix \ref{appendix:mf}) yielding the MF Hamiltonian,
\begin{equation}
    \begin{aligned}
    H_{MF}&= \displaystyle\sum_{i,j,\alpha,\sigma}\left(
    t_{ij}^{\alpha,\alpha'}c^{\alpha\dagger}_{i,\sigma}c^{\alpha'}_{j,\sigma}
    + h.c.\right)\\
    &-U\displaystyle\sum_{j,\alpha}\rho^\alpha_\uparrow
    c^{\alpha\dagger}_{j,\downarrow}c^{\alpha}_{j,\downarrow} +
    \rho^\alpha_\downarrow c^{\alpha\dagger}_{j,\uparrow}
    c^{\alpha}_{j,\uparrow}\\
    &-\displaystyle\sum_{j,\alpha}\Delta^\alpha
    c^{\alpha\dagger}_{j,\downarrow}c^{\alpha\dagger}_{j,\uparrow} +
    \Delta^{\alpha*}c^{\alpha}_{j,\uparrow}c^{\alpha}_{j,\downarrow}\\
    &-\mu\displaystyle\sum_{j,\alpha,\sigma}
    c^{\alpha\dagger}_{j,\sigma}c^{\alpha}_{j,\sigma}\\
    &+L\displaystyle\sum_\alpha
    U\rho^\alpha_\uparrow\rho^\alpha_\downarrow +
    \abs{\Delta^\alpha}^2/U.
    \end{aligned}
\end{equation}
The order parameter, $\Delta^\alpha/U = \langle
c^{\alpha}_{j,\uparrow}c^{\alpha}_{j,\downarrow}\rangle$, is complex,
in general, and sublattice-dependent. The average filling on
sublattice $\alpha$ is $\rho_{\uparrow (\downarrow)}^\alpha = \langle
c^{\alpha\dagger}_{j,\uparrow(\downarrow)}
c^{\alpha}_{j,\uparrow(\downarrow)}\rangle$. Fourier transforming, we
define the Nambu spinor, $\Psi_k^\dagger = \begin{pmatrix}
  c_{k\uparrow}^{A\dagger} &c_{k\uparrow}^{B\dagger}
  &c_{k\uparrow}^{C\dagger} &c_{-k\downarrow}^A &c_{-k\downarrow}^B
  &c_{-k\downarrow}^C \end{pmatrix}$ and write the Bogoliubov-de
Gennes Hamiltonian
\begin{equation}
\begin{aligned}
H_{MF}(\Phi) &= \displaystyle\sum_k \Psi^\dagger_k
\mathcal{M}_k(\Phi)\Psi_k \\
&+L\displaystyle\sum_\alpha\left(U\rho^\alpha_\uparrow\rho^\alpha_\downarrow
+\abs{\Delta^\alpha}^2/U-U\rho^\alpha_\downarrow -\mu\right),
\label{eq:hamBdG}
\end{aligned}    
\end{equation}
where the momentum $k$ is summed over the Brillouin zone (BZ) and
$\mathcal{M}_k(\Phi)$ is a $6 \times 6$ Hermitian matrix for the
three-band system (Appendix \ref{appendix:mf}). The phase twist $\Phi$
enters the expression through the inter-cell hopping terms.
Diagonalizing the MF Hamiltonian, Eq.(\ref{eq:hamBdG}), and solving
the self-consistent equations for the order parameters and
site-dependent fillings, we find the ground state energy
$E_{GS}(\Phi)$ and obtain the superfluid density $D_s$.

We emphasize that the full multi-band MF is crucial to describe
accurately the behavior of multi-band systems.\cite{chan2022pairing}
This means that, in the most general form, (a) the order parameters
are sublattice-dependent and complex, and (b) the sublattice-dependent
filling must be taken into account as a mean-field parameter. Without
these two ingredients, the agreement between MF and exact calculations
deteriorates, as shown in Appendix \ref{appendix:MFwithoutrhos}.

In what follows, all energies are measured in terms of our energy
scale $\abs{t_3}$, the hopping parameter between A and C sublattices
on the same unit cell.
\subsection{Flat Band Construction}
We construct the flat band lattices using the method of
Ref.[\onlinecite{maimaiti2019universal}] which we outline in Appendix
\ref{appendix:fbh}. With a choice of the CLS localized on two adjacent
unit cells in a quasi one-dimensional three-band lattice, we can
obtain lattices with a flat band as the lowest energy state. Here, we
represent the (not normalized) CLS wavefunctions on two neighboring
unit cells in their most general form in real space:
\begin{equation}
    \ket{\Psi_1} = 
    \begin{pmatrix} a \\
    be^{i\beta}\\
    ce^{i\gamma}
    \end{pmatrix}, \ket{\Psi_2} = 
    \begin{pmatrix} xe^{i\chi} \\
    ye^{i\tau}\\
    ze^{i\zeta}
    \end{pmatrix},
    \label{eq:clsvecs}
\end{equation}
where each element is the probability amplitude on the
sublattices. We can then compute the winding number, $W$,
of the flat band, through 
\begin{equation}
\label{eq:winding}
    \begin{aligned}
    W&=\frac{i}{\pi}\int_0^{2\pi}dk\,\bra{\Psi_k}\ket{\partial_k\Psi_k} \\
    &=\frac{1}{2\pi}\int_0^{2\pi}dk(1+\frac{x^2+y^2+z^2-a^2-b^2-c^2}{2\cos(k)
      +a^2+b^2+c^2+x^2+y^2+z^2}) \\
    &=1+\frac{x^2+y^2+z^2-a^2-b^2-c^2}{\sqrt{(a^2+b^2+c^2 +
        x^2+y^2+z^2)^2-4}}
    \end{aligned},
\end{equation}
where $|\Psi_k\rangle$ is the normalized Bloch state corresponding to
the CLS (Appendix \ref{appendix:fbh}).  With the condition in
Eq.(\ref{eq:winding}), we have the freedom to construct lattices with
winding numbers of our choice (Appendix
\ref{appendix:OtherLattices}). We write the kinetic energy part of the
Hamiltonian as $H_{KE}$, and $H_0$ ($H_1$) is the intra(inter)-cell
kinetic energy (see Fig. \ref{fig:lattice}),
\begin{equation}
  \label{eq:H0}
    H_0 = 
    \begin{pmatrix} 0 & t_1 & t_3 \\
    t_1 & 0 & t_2\\
    t_3 & t_2 & 0
    \end{pmatrix},
\end{equation}
\begin{equation}
\label{eq:H1}
    H_1 = 
    \begin{pmatrix} t_a & t_b & t_c \\
    t_d & t_e & t_f\\
    t_g & t_h & t_i
    \end{pmatrix}, \\
\end{equation}
\begin{equation}
    H_{KE} = \begin{pmatrix}
    H_0 & H_1 & 0 & \cdots & 0 & H_1^\dagger \\
    H_1^\dagger & H_0 & H_1 & 0 & \cdots & 0 \\
    0 & H_1^\dagger & H_0 & H_1 & \ddots & \\
    \vdots & & & \ddots
    \end{pmatrix}.
\end{equation}
Further details are outlined in Appendix \ref{appendix:fbh}. In the
examples we cover, we only consider (positive and negative) real
values for the probability amplitudes in Eq.(\ref{eq:clsvecs}).

\section{Topological Flat bands: $W\neq 0$}\label{results:nonzero_winding}

In this section, we focus on topological flat bands with nonzero
winding number, $W\neq0$. We propose two lattices, lattice $\cal A$
and lattice $\cal B$, both of which have a fixed flat band energy
$E_{FB}$ but variable dispersive bands, allowing us to tune the band
gap.

Here, the properties of the system when there is a gap between the
flat lowest energy band and the dispersive band above it are analyzed,
and distinguished from the effects when these two bands touch. In
addition, we study at fixed gap and $W$, the effect of asymmetry in
the populations of the two unit cells of the CLS, and the effect of
changing $W$.

\begin{figure}[h!]
  \includegraphics[width=8.6cm]{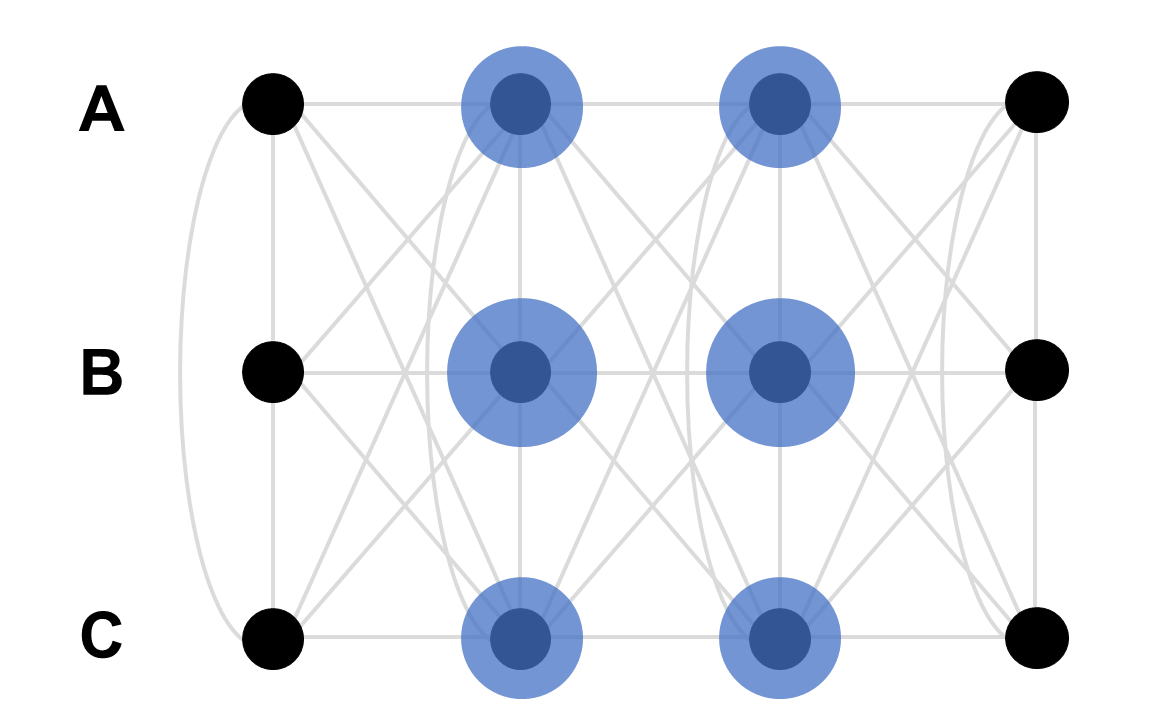}\\
  \includegraphics[width=8.6cm]{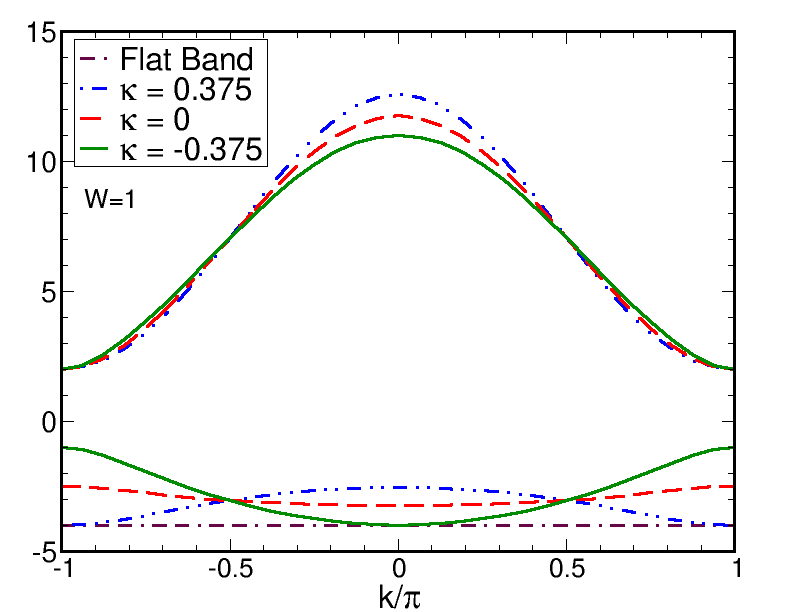}
 \caption{(Color online) Top: Compact localized states (CLS) for
   Lattice $\mathcal{A}$, Eq.(\ref{eq:CLS_A}), symmetric on both unit
   cells. The area of the blue discs is proportional to the density on
   the site. Bottom: Band structure for $\kappa=0.375$ (gapless at
   $k=\pi$), $\kappa=0$ ($E_{gap}$ $\approx0.7625$), $\kappa=-0.375$
   (gapless at $k=0$). The winding number is $W=1$.}
 \label{fig:CLS-A}
\end{figure}

We first choose a CLS with symmetric but nonuniform populations on the
two unit cells. We refer to this as lattice ${\cal A}$ and its CLS is
given by,
\begin{equation}
\label{eq:CLS_A}
    \ket{\Psi_1} = 
    \begin{pmatrix} \sqrt{2} \\
    -\sqrt{3}\\
    -\sqrt{2}
    \end{pmatrix}, \ket{\Psi_2} = 
    \begin{pmatrix} \sqrt{2} \\
    \sqrt{3}\\
    -\sqrt{2}
    \end{pmatrix},
\end{equation}
where $\ket{\Psi_1}$ and $\ket{\Psi_2}$ are the states on the two
neighboring unit cells where the CLS is found. The populations on the
two unit cells are the same, but within a unit cell, sublattice B has
a higher population than the equal populations of sublattices A and C,
shown in Fig. \ref{fig:CLS-A}. Using Eq.(\ref{eq:winding}), we show
that this choice yields a winding of $W=1$.  The hopping parameters
for this CLS are $t_1=t_2=\sqrt{7}$, $t_3=1$, $t_a = t_i =
\frac{1}{8}+\kappa$, $t_b = t_f = \frac{1}{2}(\sqrt{7}-\sqrt{6})$,
$t_c=\frac{1}{8}(13-2\sqrt{42})+\kappa$, $t_d = t_h =
\frac{1}{2}(\sqrt{7}+\sqrt{6})$, $t_e = 2$, and $t_g =
\frac{1}{8}(13+2\sqrt{42})+\kappa$ (Appendix \ref{appendix:fbh}). The
free parameter $\kappa$ is used to control the band gap, as shown in
Fig.\ref{fig:CLS-A}.  Fourier transforming $H_{KE}$, we obtain the
eigenvalues, $\lambda_i$, which describe the band structure in
Eq.(\ref{eq:eigenvalues}).

The flat band energy is fixed at $E_{FB}=-4$, independent of
$\kappa$. For the gapped case we put $\kappa=0$, while for the gapless
cases, $\kappa=-0.375$ which has the lower two bands touching at $k=0$
($\frac{\partial^2\lambda_2}{\partial k^2}\vert_{k=0}=\frac{4}{5}$)
and $\kappa = 0.375$ is gapless at $k=\pi$
($\frac{\partial^2\lambda_2}{\partial k^2}\vert_{k=\pi}=\frac{3}{2}$),
depicted in Fig. \ref{fig:CLS-A}.

\widetext
\begin{equation}
    \begin{aligned}
      &\text{\small$\lambda_1=-4$}\\
      &\text{\small$\lambda_2 = \tfrac{1}{8} \left( 16 + 18 \cos
        \left(k\right)+16\kappa\cos(k)-\sqrt{6\cdot \left( 299+272
          \cos\left(k\right)+27\cos\left(2k\right)\right)+2 \left(
          128\kappa^{2}\cos^{2}(k)-128\kappa\cos(k)+160\kappa
          \cos^{2}(k) \right)}\right)$}\\
      &\text{\small$\lambda_3=\tfrac{1}{8} \left(16+18 \cos
        \left(k\right)+16\kappa\cos(k)+\sqrt{6\cdot \left( 299 +
          272\cos\left(k\right)+27\cos\left(2k\right)\right) +
          2\left(128\kappa^{2}\cos^{2}(k)-128\kappa\cos(k) +
          160\kappa\cos^{2}(k)\right)}\right)$}
    \end{aligned}
    \label{eq:eigenvalues}
\end{equation}
\twocolumngrid

\subsection{Gapped Case}

\begin{figure}[h!]
 \includegraphics[width=8.6cm]{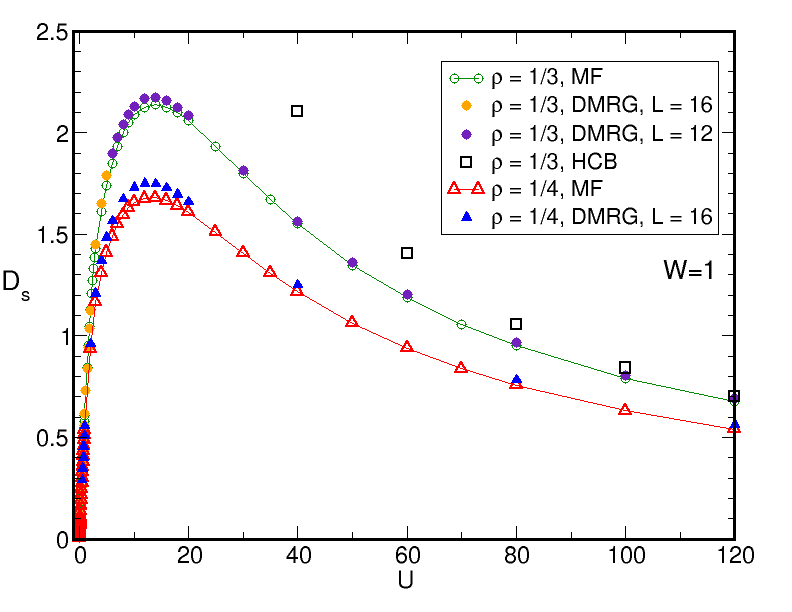}
 \caption{(Color online) Lattice $\mathcal{A}$ ($W=1$) gapped case
   ($\kappa=0$): the superfluid density, $D_s$, computed with DMRG and
   MF for $\rho=1/4$ and $\rho=1/3$. DMRG and MF agree for the entire
   range of $U$ for both fillings, approaching the hard-core boson
   (HCB) limit at strong interaction.}
 \label{fig:DsvsU-CLSA}
\end{figure}

We study first the gapped case with $E_{gap}\approx0.7625$
($\kappa=0$). Figure \ref{fig:DsvsU-CLSA} shows the superfluid
density, $D_s$, versus the interaction strength, $U$, computed with
DMRG and MF for two fillings, $\rho=1/4$ and $\rho=1/3$, where
$\rho=1/3$ gives a half-filled flat band. The agreement is excellent
between DMRG and MF for both densities and over the entire wide range
of $U$ values. In addition, the hard-core boson (HCB)
approximation\cite{emery1976theory} agrees with exact and MF values of
$D_s$ at very large $U$, when the Cooper pairs are tightly bound. In
this limit, the transport of effective hard-core bosons is governed by
a dispersive model with repulsive nearest-neighbour
interaction.\cite{emery1976theory} At low $U$, $D_s$ rises linearly
with $U$, as has been established for isolated flat bands. The slope
of $D_s$ against $U$ is $0.538$ for $\rho=1/4$ and $0.618$ for
$\rho=1/3$. In addition to $D_s$, the order parameters and sublattice
fillings also show excellent agreement between MF and DMRG results at
$\Phi=0$, as we show in Appendix \ref{appendix:Bands_Deltas}.

\begin{figure}[h!]
 \includegraphics[width=8cm]{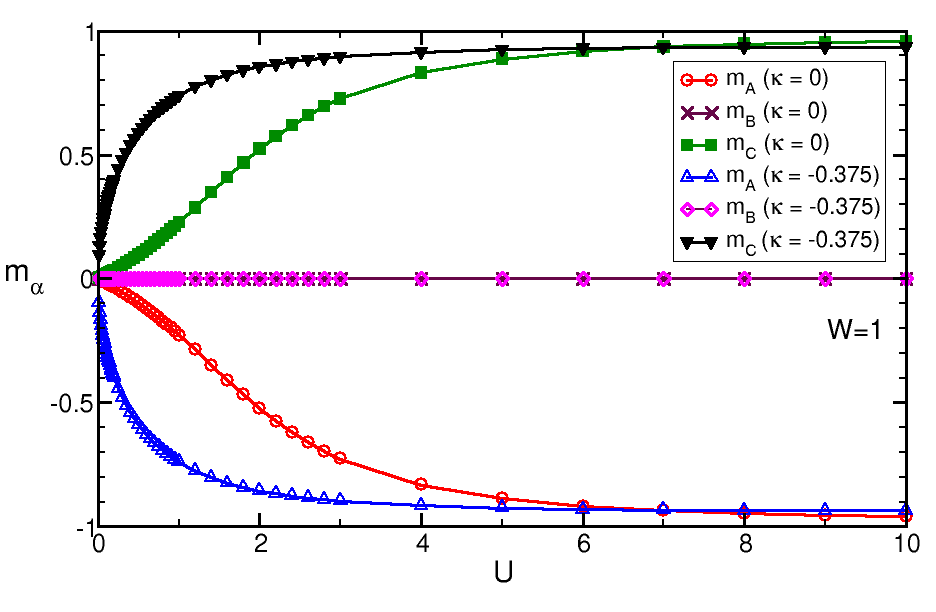}
 \caption{(Color online) Lattice $\mathcal{A}$ ($W=1$):
   Proportionality factor $m_\alpha(U)$,
   Eq.(\ref{eq:phase-condition-CLS-A}), versus $U$ for the gapped and
   gapless band structures of lattice ${\mathcal A}$.  At large $U$,
   $m_\alpha \to 1$ and the $U$-dependence saturates. The filling is
   $\rho=1/3$. The same behavior is observed at other fillings.}
 \label{fig:Phases-CLSA}
\end{figure}
Furthermore, while the above features are similar to ones we have
observed previously\cite{chan2022pairing}, this system exhibits a
property not encountered before. In the case of the Creutz flat band
system we found that $\Delta^A=\Delta^B$ and both can be taken to be
real; for the sawtooth system, we found that $\Delta^A\neq \Delta^B$,
so that in general one can be taken real but the other
complex.\cite{chan2022pairing} In addition, we found for the sawtooth
lattice, that when a phase gradient is applied, the phase difference
between the order parameters on the two sublattices was constant and
equal to the phase gradient. In the present case, we see from
Eq.(\ref{eq:CLS_A}) that sublattices A and C have equal fillings, and
MF shows that they also have the same magnitude of the complex order
parameter. However, they do not have the same phase when a phase twist
is applied, $\Phi\neq 0$. By doing a global gauge transformation, the
phase of $\Delta^B=\abs{\Delta^B}e^{i\theta_B}$ can be put to $0$. We
then find that for $\Delta^A=\abs{\Delta^A}e^{i\theta_A(U)}$ and
$\Delta^C=\abs{\Delta^C}e^{i\theta_C(U)}$, the magnitudes are equal,
$\abs{\Delta^A}=\abs{\Delta^C}\neq \abs{\Delta^B}$, and when
$\theta_B=0$, $\theta_A=-\theta_C$. Therefore, while sublattices $A$
and $C$ appear to be equivalent, the phases of the order parameters
are opposite in sign. This can be proved as follows. For the lattice
we are considering, both matrices $H_0$ and $H_1$,
Eqs.(\ref{eq:H0},\ref{eq:H1}), are real matrices, so that one can
easily prove that for each sublattice,
$\Delta^{\alpha}(-\phi)=\Delta^{\alpha *}(\phi)$, a situation similar
to a system invariant under time reversal. In addition, the structure
of $H_0$ and $H_1$ is such that exchanging sublattices $A$ and $C$
amounts to a parity-like symmetry, i.e. changing $i\rightarrow -i$ and
$\phi\rightarrow-\phi$, in the Hamiltonian. When combined with the
preceding properties, this allows us to show that
$\Delta^{A*}(\phi)=\Delta^C(\phi)$ and thereby that
$|\Delta^A|=|\Delta^C|$ and $\theta^A=-\theta^C$.

We also find here that, contrary to the sawtooth case, the phases of
the order parameters are not constant but are $U$-dependent. At
constant $U$, they are proportional to the phase gradient,
\begin{equation}
    \theta_\alpha(U)=m_\alpha(U)\phi,
    \label{eq:phase-condition-CLS-A}
\end{equation}
where $m_\alpha(U)$ is a $U$-dependent proportionality factor. This is
shown in Fig. \ref{fig:Phases-CLSA} where we see that at large $U$,
$m_\alpha(U)\to 1$, exhibiting in that saturated limit behavior
similar to that of the sawtooth lattice where the phases are not
$U$-dependent. Note that $m_\alpha(U)$ changes very rapidly for small
$U$ where $D_s$ is linear in $U$. This emphasizes, yet again, the
importance of including three distinct sublattice-dependent, complex
order parameters (in addition to the sublattice-dependent fillings)
when describing these systems with mean field methods.

At this point, we have examined the properties of a specific choice
for the CLS, lattice ${\cal A}$, Eq.(\ref{eq:CLS_A}). However, as
mentioned above, there is a lot of freedom in $|\Psi_1\rangle$ and
$|\Psi_2\rangle$ while keeping constant $E_{FB}$, $W$, and the gap. An
interesting case to consider is a CLS with uniform site densities on
all sublattices and both unit cells. We call this
lattice ${\mathcal B}$, and its CLS is given by,

\begin{equation}
    \ket{\Psi_1} = \begin{pmatrix}
    1 \\ -1 \\ 1
    \end{pmatrix}, \quad \ket{\Psi_2} = \begin{pmatrix}
    1 \\ 1 \\ 1
    \end{pmatrix}.
    \label{eq:latticeB}
\end{equation}

This choice has $W=1$ and intra- and inter-cell hopping Hamiltonians,
\begin{equation}
  \label{eq:latticeBham}
    H_0 = 
    \begin{pmatrix} 0 & -1 & -1 \\
    -1 & 0 & 1\\
    -1 & 1 & 0
    \end{pmatrix}, H_1 = 
    \begin{pmatrix} \nu & -1 &  -1-\nu\\
    0& 1 & 1 \\
    -\nu & 0& \nu
    \end{pmatrix},
\end{equation}
where $\nu$ is a parameter used to tune the gap. The eigenvalues
describing the three bands are, \widetext
\begin{equation}
    \label{eq:latticeBeigens}
    \begin{aligned}
        \lambda_1 &= -2 \\
        \lambda_2 &= 1+\cos(k)+2\nu\cos(k)-\sqrt{4\nu^2\cos^2(k) +
          \cos^2(k)+4\cos(k)+3} \\ 
        \lambda_3 &= 1+\cos(k)+2\nu\cos(k)+\sqrt{4\nu^2\cos^2(k) +
          \cos^2(k)+4\cos(k)+3}
    \end{aligned}
\end{equation}
\twocolumngrid These bands touch at $k=0$ for $\nu=-0.5$
($\frac{\partial^2\lambda_2}{\partial k^2}\vert_{k=0}=\frac{4}{3}$)
and $k=\pi$ for $\nu=0.5$ ($\frac{\partial^2\lambda_2}{\partial
  k^2}\vert_{k=\pi}=2$) with a flat band energy $E_{FB}=-2$,
independent of $\nu$. For the gapped case, we choose $\nu=-1/12$ for a
band gap of $E_{gap}=1$.

\begin{figure}[h!]
 \includegraphics[width=8cm]{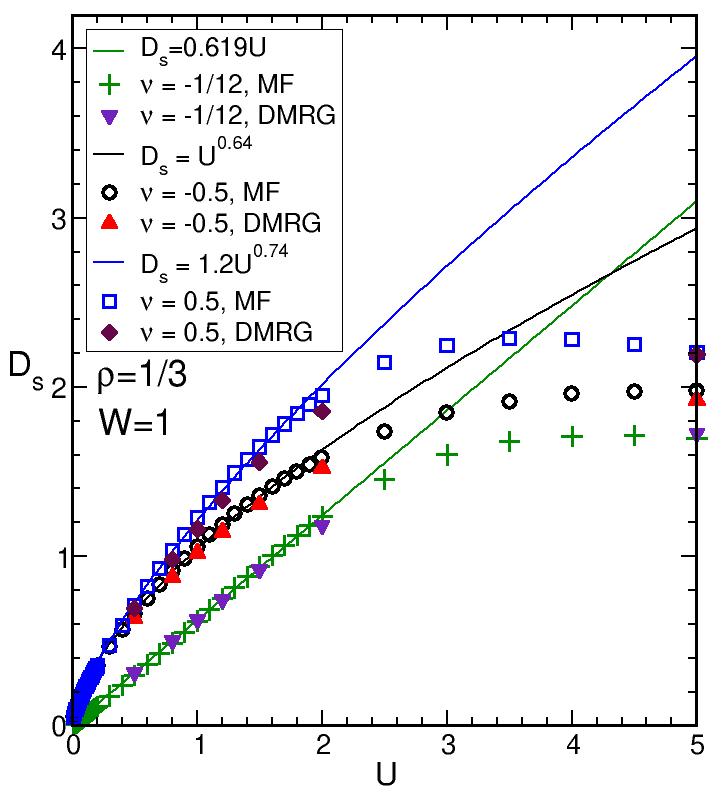}
 \caption{(Color online) Lattice $\mathcal{B}$ ($W=1$): $D_s$ vs $U$
   with $\nu=-1/12$ ($E_{gap}=1$), $\nu = -0.5$ (gapless at $k=0$),
   $\nu=0.5$ (gapless at $k=\pi)$. We observe a linear relation for
   the gapped case and $D_s\sim U^\varphi$ with $\varphi<1$ for the
   touching bands. The filling is $\rho=1/3$.}
 \label{fig:DsvsU-CLS-B}
\end{figure}

\begin{figure}[h]
 \includegraphics[width=8.6cm]{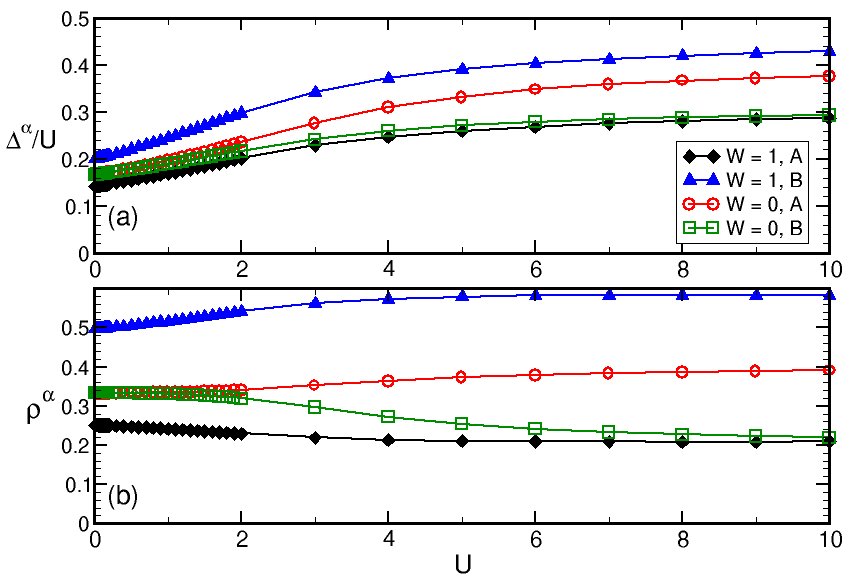}
 \caption{(Color online) MF order parameters and fillings on
   sublattices A and B for isolated topological (lattice $\mathcal{B},
   W=1$) and non-topological (lattice $\mathcal{C}, W=0$) flat bands
   with $E_{gap}=1$ and equal filling on all sites of CLS. Lattice
   $\mathcal{B}$ with $W=1$ has $\Delta^A\neq\Delta^B$ and
   $\rho^A\neq\rho^B$ for any finite $U$. For lattice ${\cal C}$
   ($W=0$), the sublattice dependent order parameters and fillings go
   smoothly to equal values as $U\rightarrow0$. In both cases,
   $\abs{\Delta^A}=\abs{\Delta^C}$ and $\rho^A=\rho^C$, but the order
   parameters on all three sublattices have different phases at
   $\phi\neq0$. When $W=0$, $\Delta^B$ ($\rho^B$) is smaller than
   $\Delta^A=\Delta^C$ ($\rho^A=\rho^C$), in contrast to the
   topological $W=1$ cases. The filling is $\rho=1/3$.}
 \label{fig:DeltasDensities-eqCLS}
\end{figure}

Figure \ref{fig:DsvsU-CLS-B} shows that, in the gapped case
($\nu=-1/12$), $D_s$ is again linear in $U$ for lattice ${\cal B}$ and
small $U$ with a slope of $0.619$. Naively, one might expect the
uniformity of site fillings to persist when $U\neq0$. On the contrary,
we find that for any finite $U$, the uniform density of the CLS of
lattice ${\cal B}$ (Eq.(\ref{eq:latticeB})) is broken and the
sublattices again become distinct. This is seen clearly in
Fig. \ref{fig:DeltasDensities-eqCLS} where (for $W=1$, the case we
examine here) $\Delta^A/U \neq \Delta^B/U$ and $\rho^A \neq \rho^B$
for all $U>0$. The phases of the order parameters, $\Delta^A$ and
$\Delta^C$, behave in the same way as for lattice ${\cal A}$,
i.e. they are equal and opposite in sign when the global gauge is
fixed so that $\Delta^B$ is real.

For lattices $\cal A$ and $\cal B$ with symmetric CLS on both unit
cells, we obtain approximately equal slopes of $D_s$ against $U$ for
the gapped case (Fig. \ref{fig:DsvsU-CLS-B} and
Fig. \ref{fig:CLS-A_lowUcomparison}) despite the differences in band
gap, hopping potentials and flat band energy. One might argue that
this should be obvious, as transport is dominated by the flat band and
they have equal winding numbers. However, with the freedom of
constructing asymmetric CLS with the same $W$, $E_{gap}$, and $E_{FB}$
as the symmetric ones, we demonstrate the dominant effect of the CLS
symmetry on SC properties. In Appendix \ref{appendix:OtherLattices},
we show in detail that a symmetric CLS on the flat band is the most
favourable in terms of optimising the SC, for $W=1$. Maintaining a
winding of 1, $E_{FB}=-4$, and $E_{gap}\approx 0.7625$, we find that
the more asymmetric the CLS is, the slower $D_s$ increases with
$U$. Very interestingly, as the slope of the linear part of $D_s(U)$
(which we denote by ${\cal S}$) decreases due to the asymmetry, the
value of the integral over the BZ of the quantum metric remains rather
constant: The slope is much more sensitive to the CLS than to the
quantum metric (Appendix \ref{appendix:OtherLattices}). Additionally,
the site fillings, order parameters, and band structures do not vary
much across the cases considered, despite the significant difference
in $D_s$.

As mentioned above, the winding number of the flat band can be
tuned. To illustrate this, we constructed Hamiltonians with $W=1/2$,
for several CLS configurations, and performed a similar study as $W=1$
(Lattices $\mathcal{D}1, \mathcal{D}2, \mathcal{D}3$ in Appendix
\ref{appendix:OtherLattices}). When $W=1/2$, Eq.(\ref{eq:winding})
dictates that the CLS must be asymmetric. Qualitatively, the behavior
of $D_s$ is similar to the $W=1$ case in that it exhibits a linear
part at low $U$. $\Delta^A$, $\Delta^B$, $\Delta^C$ and their phases
are unequal for all sublattices and dependent on $U$. As we reduce the
filling on one unit cell of the CLS, increasing the asymmetry, we find
the slope, $\cal S$, of $D_s(U)$ decreases. In other words, for the
same sublattice, the occupation must be comparable on both unit cells
of the CLS to increase $D_s$.

We thus conclude this subsection by stating that to optimize $D_s$ on
the isolated topological flat band, one should identify the case with
the most symmetric CLS. Specifically, when $W=1$, the occupation on
the optimized CLS will be truly symmetric.

\subsection{Gapless Cases}

It has been argued that non-isolated flat bands may be beneficial to
SC\cite{huhtinen2022revisiting,iskin2019origin,wu2021superfluid,julku2016geometric}
with $D_s$ increasing as $U{\rm ln}({\rm const.}/U)$ for small $U$,
i.e. faster than linear. Here, we study this situation where a
dispersive band touches the flat band below it and what effect it has
on SC. To this end, we consider our lattices ${\cal A}$ and ${\cal B}$
both with $W=1$. The former has a symmetric CLS but with site
densities which are not uniform,
Eqs.(\ref{eq:CLS_A},\ref{eq:eigenvalues}); the latter has a CLS with
uniform site densities,
Eqs.(\ref{eq:latticeB},\ref{eq:latticeBham},\ref{eq:latticeBeigens}).
For lattice ${\cal A}$ we take $\kappa=-0.375$ (bands touching at
$k=0$) and $\kappa =0.375$ (bands touching at $k=\pi$; for lattice
${\cal B}$, we take $\nu=-1/2$, (bands touching at $k=0$) and
$\nu=1/2$ (bands touching at $k=\pi$).

\begin{figure}[h!]
 \includegraphics[width=7.8cm]{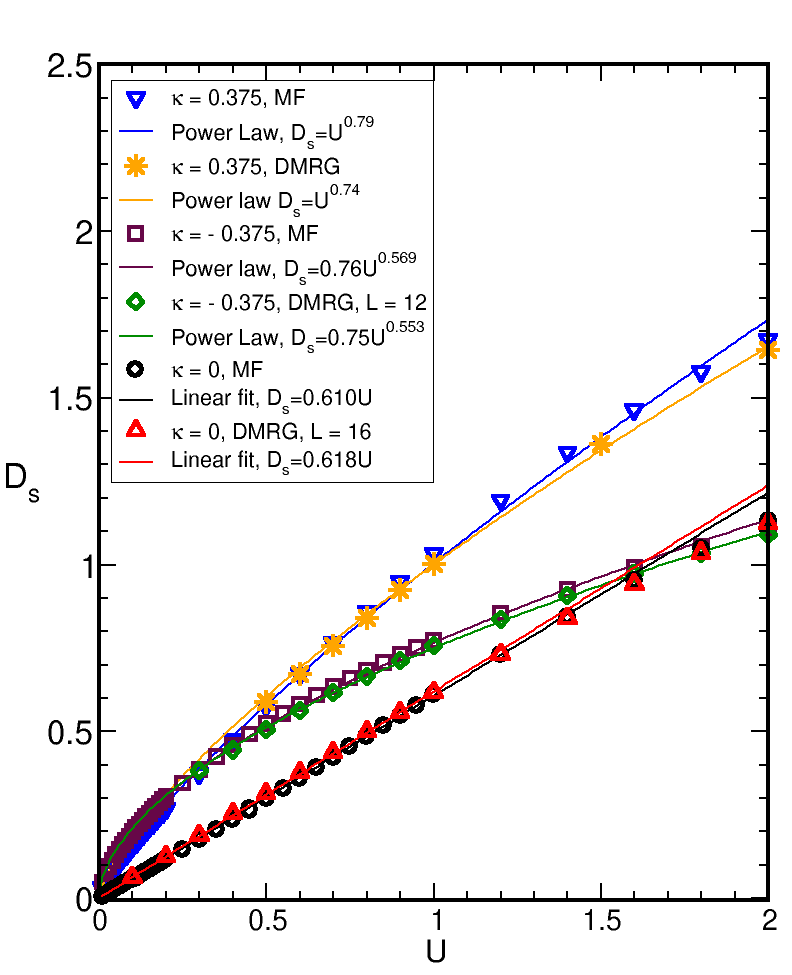}\\
   \includegraphics[width=8.8cm]{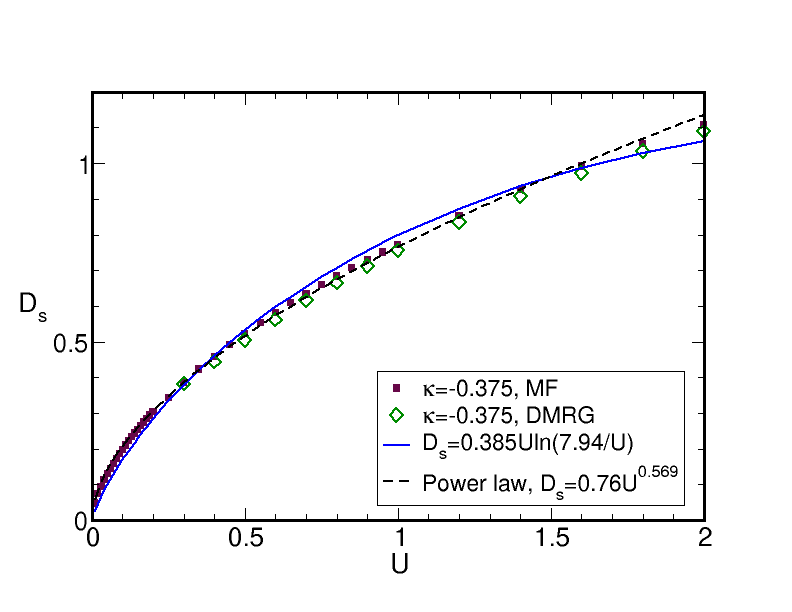}
 \caption{(Color online) Top: The low $U$ behavior of $D_s$ for
   Lattice $\mathcal{A}$ for $\rho=1/3$, with $\kappa=-0.375$
   (touching at $k=0$), $\kappa=0$ (gapped), and $\kappa= 0.375$
   (touching at $k=\pi$) show that touching bands can drastically
   improve superconductivity at weak attraction. There is a clear
   power law dependence, with power exponent less than $1$ for both
   $\kappa=-0.375$ and $\kappa=0.375$. Bottom: comparing power
   law, $D_s \propto U^\varphi$, and $D_s \propto U{\rm ln}({\rm const.}/U)$
   fits. The data are inconsistent with a $U{\rm ln}({\rm const.}/U)$
   behavior.}
 \label{fig:CLS-A_lowUcomparison}
\end{figure}

In Figs. \ref{fig:DsvsU-CLS-B} and \ref{fig:CLS-A_lowUcomparison} we
exhibit the behavior of $D_s$ at low values of $U$ for both these
gapless systems and we also include the corresponding linear gapped
case for comparison. We see that for all gapless cases (bands touching
at $k=0,\pi$), a power law fit, $D_s \propto U^\varphi$ with
$\varphi<1$, describes the dependence very well for both DMRG and
MF. This means that for small $U$, $D_s$ increases faster with $U$
when the bands touch than when there is a gap where the behavior is
linear; this favors SC because the carrier density is higher at low
$U$. However, in both gapped and gapless cases, the $\Delta$s increase
linearly with $U$. In the quasi one-dimensional case we examine here,
there is no finite temperature transition between SC and a normal
phase: true SC is present at $T=0$ only. If the $D_s \propto
U^\varphi$ with $\varphi<1$ and $\Delta^\alpha \propto U$ behavior
persists in higher dimensions, that would mean that $T_c$ may not be
enhanced when the bands touch, since $\Delta^\alpha\propto U$, even
though the carrier density itself is enhanced. In the bottom panel of
Fig. \ref{fig:CLS-A_lowUcomparison}, we compare the quality of fits of
the power law and the $U{\rm ln}({\rm const.}/U)$. We find the power
law to be in much better agreement with MF and DMRG and over a wider
range of $U$; we therefore argue that the power law is more
appropriate than the logarithmic form to describe $D_s$. We point out
two main differences between our systems and those discussed in
Refs.[\onlinecite{huhtinen2022revisiting,iskin2019origin,wu2021superfluid,julku2016geometric}]
where the logarithmic behavior is observed: (a) our systems are quasi
one-dimensional where as those in
Refs.[\onlinecite{huhtinen2022revisiting,iskin2019origin,wu2021superfluid,julku2016geometric}]
are two-dimensional, and (b) in our systems, unlike the
two-dimensional ones, when the gap between the flat band and the
second band shrinks, the CLS remains unchanged and, consequently, the
quantum metric and its BZ integral remain constant. In the
two-dimensional systems mentioned, the logarithmic behavior of $D_s$
has been attributed to a logarithmic divergence of the BZ integral of
the diagonal quantum
metric.\cite{huhtinen2022revisiting,iskin2019origin,wu2021superfluid,julku2016geometric}
In our case, as we mentioned, the CLS remains unchanged as the gap
shrinks and the bands touch, and consquently no such divergence
occurs.

\begin{figure}[h!]
  \includegraphics[width=8.5cm]{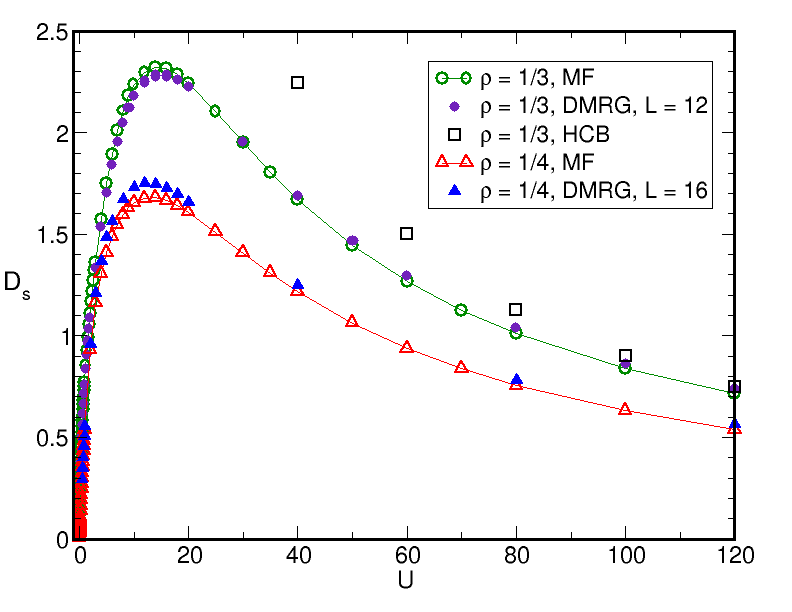}
 \caption{(Color online) Lattice $\mathcal{A}$ ($W=1$) gapless at $k=0$
   ($\kappa=-0.375$). The superfluid density obtained with MF and
   exact DMRG for two fillings, $\rho=1/4$ and $\rho=1/3$. Even with
   touching bands, our MF agrees very well with DMRG.}
 \label{fig:gaplessCLSW1}
\end{figure}

We note that the power law exponents are larger
when the curvature of the dispersive band is larger which, for these
two lattices, happens at $k=\pi$. In Fig. \ref{fig:gaplessCLSW1} we
show $D_s$ over a very wide range of $U$ for two densities when the
bands touch at $k=0$. We see, again, that agreement between our
multi-band MF and exact DMRG is excellent over the entire range of
$U$, as it was in the gapped case. The order parameters are shown in
Appendix \ref{appendix:Bands_Deltas}. In addition, we use this case to
illustrate the inaccuracy of MF when the site densities are not
included as variational MF parameters (Appendix
\ref{appendix:MFwithoutrhos}).

\begin{figure}[h!]
 \includegraphics[width=8.6cm]{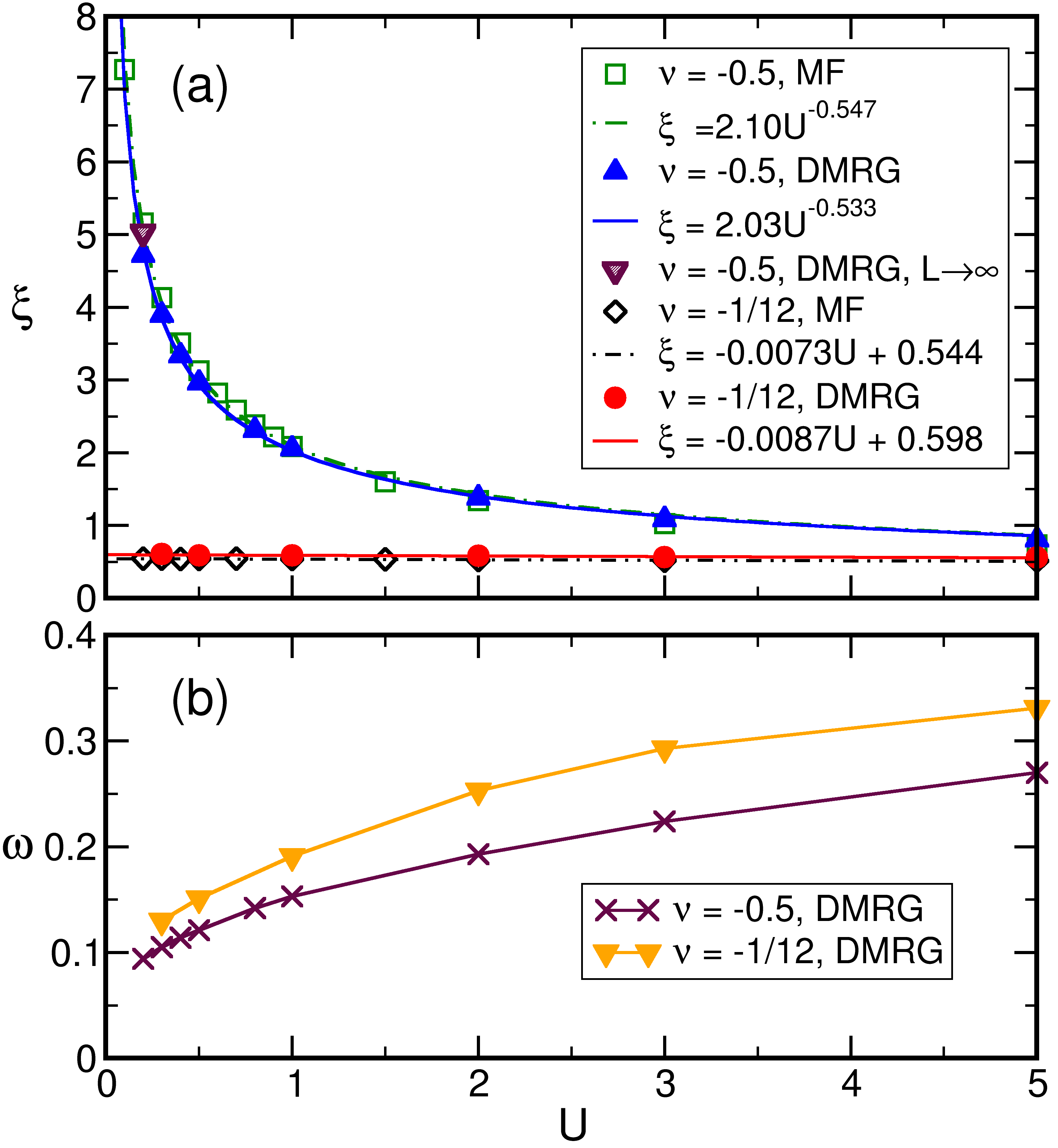}
 \caption{(Color online) Lattice $\cal B$. (a) Correlation length
   $\xi$ with fitted functions for both MF and DMRG ($L=100$). For
   $U=0.2$, we include the extrapolated value $\xi(L\to\infty)$ for
   touching bands, where finite size effects result in a slightly
   increased discrepancy between MF and DMRG at low $U$. (b) Pair
   Green's function power law decay exponent, $\omega$ obtained from
   DMRG, as a function of $U$ for the cases where $E_{gap}=1,
   \nu=-1/12$ and gapless at $k=0$, $\nu=-0.5$. The filling is $\rho =
   1/3$.}
 \label{fig:corrlengthCLSB}
\end{figure}

As a result of pairing, the single particle Green's functions exhibit
exponential decay, while the pair Green's functions decay with a power
law since the system is SC\cite{gremaud2021pairing},
Eq.(\ref{eq:GF}). The correlation length, $\xi$, extracted from the
decay exponent of the single particle GF obtained with our MF method
agrees very well with that obtianed with DMRG, which we show for
Lattice $\cal B$ in Fig. \ref{fig:corrlengthCLSB}(a). Recalling that
the correlation length typically diverges exponentially for dispersive
bands\cite{gremaud2021pairing} as $U\rightarrow0$; we find here a
different behavior for the correlation length on the gapped and
gapless flat bands. As previously established in
Ref.[\onlinecite{chan2022pairing}] for the Creutz and sawtooth
lattices, the correlation length goes to a constant, less than one
lattice spacing for the isolated flat band as $U\rightarrow0$. We
observe here this same behavior in the gapped case. However, in the
gapless case, the single particle correlation length, $\xi$, diverges
as a power law $\xi\sim U^{-P}$, as $U\rightarrow0$, i.e. much slower
than in the dispersive case.  The power law decay of the pair Green's
function is characterized by the exponent $\omega$ which we calculate
with DMRG. We find that $\omega$ increases with $U$ and is larger in
the gapped case than in the gapless case. The lower values of $\omega$
in the gapless case is consistent with the larger values of $D_s$ than
for the isolated flat band in the same range of
$U$. (Fig. \ref{fig:DsvsU-CLS-B}). In other words, the smaller
$\omega$ is, the slower the decay of the quasi-long range order and
the larger the $D_s$. We mention that the pair Green's functions
cannot be obtained using MF.

\section{Non-topological Flat Bands: $W=0$}\label{results:zerowinding}

In this section we study pairing and superconductivity in a system
with a non-topological flat band with zero winding number, $W=0$. To
this end, we exploit the tunability of the winding number in the
approach we explained above, and tune the CLS to yield a Hamiltonian
with $W=0$ for the flat band. We found that to accomplish this, the
CLS must be localized within only one unit cell rather than two
neighboring unit cells as is the case for the topological bands
(details in Appendix \ref{appendix:fbh}). We call this non-topological
case lattice ${\cal C}$. We choose it to have equal sublattice
densities on all sites, and a flat band energy $E_{FB}=-2$. The CLS is
given by,
\begin{equation}
    \ket{\Psi_1} = \begin{pmatrix}
    1 \\ 1 \\ 1
    \end{pmatrix},
\end{equation}
the hopping terms are
\begin{equation}
\begin{aligned}
    H_0 &= 
    \begin{pmatrix} 0 & -1 & -1 \\
    -1 & 0 & -1\\
    -1 & -1 & 0
    \end{pmatrix}, \\
    H_1 &= QKQ, \\
    Q &= \frac{1}{3}\begin{pmatrix}2 & -1 & -1 \\
    -1 & 2 & -1 \\
    -1 & -1 & 2     \end{pmatrix}.
\end{aligned}
\end{equation}
$K$ in this case is an arbitrary nonzero $3 \times 3$ matrix. To have
a band gap of 1, a possible, but not unique, choice for $K$ is,
\begin{equation}
    K = \begin{pmatrix}
    0 & 0 & 0 \\
    0 & 0 & 0 \\
    2 & 0 & 0
    \end{pmatrix},
\end{equation}
which gives the eigenvalues
\begin{equation}
    \begin{aligned}
    \lambda_1 &= -2, \\
    \lambda_2&=\frac{1}{3}\left(3-2\cos(k)-\sqrt{14+2\cos(k)}\right),\\
    \lambda_3&=\frac{1}{3}\left(3-2\cos(k)+\sqrt{14+2\cos(k)}\right),
    \end{aligned}
\end{equation}
For the gapless case, we use the construction
\begin{equation}
    K = \begin{pmatrix}
    0 & 0 & 3/2+\varrho  \\
    0 & 0 & 0 \\
    3/2-\varrho & 0 & 0
    \end{pmatrix},
\end{equation}
for the bands touching at $k=0$, and 
\begin{equation}
    K = \begin{pmatrix}
    0 & 0 & -3/2+\varrho  \\
    0 & 0 & 0 \\
    -3/2-\varrho & 0 & 0
    \end{pmatrix},
\end{equation}
for the bands touching at $k=\pi$, where $\varrho$ is a parameter
which controls the upper bands and consequently, their curvature.  We
then arrive at
\begin{equation}
    \begin{aligned}
    \lambda_1 &= -2, \\
    \lambda_2&=1\mp\cos(k)-\frac{\sqrt{6}}{3}\sqrt{(3-\varrho^2)
      \cos(2k)+3+\varrho^2}, \\
    \lambda_3&=1\mp\cos(k)+\frac{\sqrt{6}}{3}\sqrt{(3-\varrho^2)
      \cos(2k)+3+\varrho^2},
    \end{aligned}
\end{equation}
for the gapless case, $-$ ($+$) for band touching at $k=0$ ($k=\pi$),
in $\lambda_2$ and $\lambda_3$, and we can calculate exactly the
curvature of the second band.

Before examining the properties arising from filling lattice $\cal C$,
we point out some apparent differences between the band structures of
the $W\neq 0$ and $W=0$ cases. For the $W\neq 0$ lattices, we had the
option of having the bands touch at $k=0$ or at $k=\pi$, but did not
have the freedom to tune the curvature at the point where they
touch. Here, we can control both the touching point and the
curvature. For a fixed $\varrho$, the band curvature is equal where
they touch --- at $k=0$ and $k=\pi$.

First, we highlight the differences between lattices $\mathcal{B}$
($W=1$ with uniform site densities in the CLS) and $\mathcal{C}$
($W=0$ with uniform site densities in the CLS), in terms of sublattice
equivalence. For comparison, we use the isolated bands case.  Figure
\ref{fig:DeltasDensities-eqCLS} shows that for any nonzero $U$, no
matter how small, the order parameters and site densities are no
longer uniform in the $W=1$ case despite supporting a uniform CLS. On
the other hand, for $W=0$, we see that as $U\to 0$, the order
parameters on the two sublattices approach each other and merge at low
$U$. The same behavior is observed for the site
densities. Furthermore, while the magnitude of pairing order parameter
on sublattices A and C are equal, their phases for $\Phi\neq0$ are
not. We thus reiterate that even when the CLS is uniform, it is
prudent always to consider independent complex order parameters and
site densities as variational parameters when applying the MF method.

Even though the pairing parameters, $\Delta^{A,B}/U$, are finite as
$U\to 0$ in the isolated flat band case
(Fig. \ref{fig:DeltasDensities-eqCLS}), indicating robust pairing for
any $U$, the superfluid density itself is suppressed: It decays as a
power as $U\to 0$. For $E_{gap}=1$, MF yields $D_s=0.028U^{2.32}$ This
power law decay of $D_s$ can be understood through projecting the MF
onto the flat band and examining the terms that
contribute.\cite{chan2022pairing} The leading term proportional to $U$
vanishes for $W=0$ under this construction, which has a CLS localized
on one unit cell. As a result, the first nonzero term is of a higher
order. DMRG convergence becomes increasingly difficult and time
consuming at low values of $U$. The flat non-topological band can
neither contribute to transport through the band curvature nor
topology, resulting in an increased number of DMRG sweeps and states
required at low $U$ where band mixing is highly suppressed.

\begin{figure}[h]
 \includegraphics[width=8.5cm]{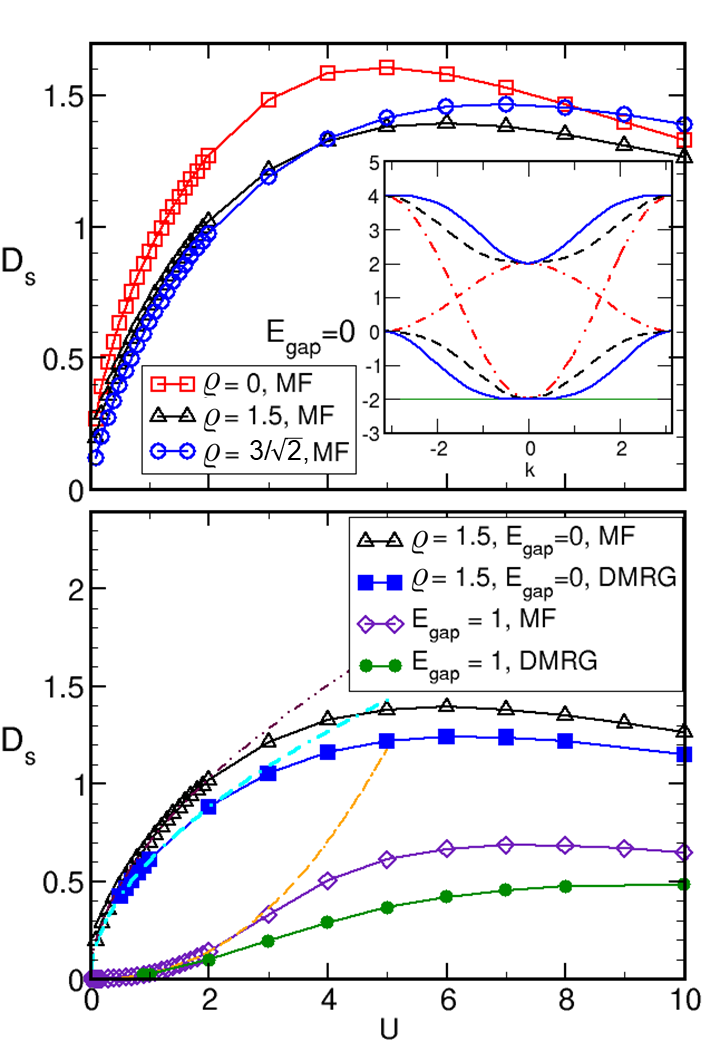}
 \caption{(Color online) $D_s$ vs $U$ for $W=0$ isolated and
   non-isolated flat band. Top: MF shows that when the bands touch,
   the curvature and optimization of $D_s(U)$ depend on the curvature
   of the second band, controlled by $\varrho$. Inset: Band structure
   of Lattice $\mathcal{C}$ for $\varrho=0,1.5,3/\sqrt{2}$ in the
   gapless case. Increasing $\varrho$ decreases the curvature of the
   second band. Bottom: Suppressed superconductivity in the gapped
   case behaves as a power law, with the fit $D_s=0.028U^{2.32}$ at
   low $U$ for MF. While DMRG and MF differ numerically, the power
   dependence of $D_s$ obtained from both methods agree (shown for
   $\varrho=1.5$). DMRG: $D_s = 0.612U^{0.53}$, MF: $D_s =
   0.7U^{0.55}$. The filling is $\rho=1/3$.}
 \label{fig:gaplessDsvsUW0}
\end{figure}

When the flat band touches the band above it, we find that $D_s$ is
strongly enhanced and grows as a power, $D_s\propto U^\varphi$ with
the exponent $\varphi < 1$,
Fig. \ref{fig:gaplessDsvsUW0}. Interestingly, this is exactly the same
behavior we found when the topological flat band touches the band
above it. In addition, as mentioned above, in this case we also have
the freedom to tune the curvature of the second band. In our
construction, the limits of the curvature are (while avoiding band
crossing) at $\varrho=3/\sqrt{2}$
($\frac{\partial^2\lambda_2}{\partial k^2}\vert_{k=0}=0$) and
$\varrho=0$ ($\frac{\partial^2\lambda_2}{\partial
  k^2}\vert_{k=0}=3$). We show in Fig. \ref{fig:gaplessDsvsUW0} three
examples of curvatures (the aforementioned and $\varrho=1.5,
\frac{\partial^2\lambda_2}{\partial k^2}\vert_{k=0}=1.5$) and their
corresponding $D_s$. Transport is dominated by the upper band in the
non-topological case and the effective mass of fermions on the upper
band decreases with increasing band curvature. However, with the bands
touching and degenerate states supported by the flat band, the
behavior of $D_s$ as $U\rightarrow 0$ is unlike the exponential decay
of a dispersive band. Consequently, increasing the curvature of the
second band decreases the power exponent, $\varphi$, with $D_s\propto
U^\varphi$, where the steepest curvature of the second band is most
beneficial towards optimizing the superfluid behavior.

In all cases, the phases of the order parameters behave similarly to
Fig. \ref{fig:Phases-CLSA}, in that they differ for all three
sublattices and are dependent on the interaction strength. We notice
that the isolated band case has significantly smaller SC density,
$D_s$, than the gapless case for the entire range of $U$, unlike the
$W\neq0$ case where the SC densities are comparable once there is band
mixing, i.e. once $U$ is of the order of the gap energy.

When the bands touch at $k=\pi$, we observe identical behavior, where
for the same $\varrho$ and upper band curvature, values of $D_s$,
order parameters and sublattice fillings are equal to when the bands
touch at $k=0$. This highlights the fact that the superfluid weight on
the gapless, non-topological flat band is only controlled by the upper
band curvature. Additionally, the correlation length on
non-topological flat bands as $U\to 0$ is identical to that in
Fig. \ref{fig:corrlengthCLSB}, with power law divergence for touching
bands and a constant, less than one lattice spacing, for isolated flat
bands.

We remark that while for $W=0$ the qualitative agreement between MF
and DMRG is excellent, the quantitative agreement for $W\neq 0$ is
much better. We believe this is due to the fact that the CLS for $W=0$
is on a single unit cell with the consequence that hopping between
unit cells, i.e. transport, requires the participation of the higher
band which is a higher order process. On the other hand, for $W\neq
0$, the CLS is spread over two unit cells to begin with, which makes
transport easier.

\section{Conclusions}\label{concs}

In this work we have extended the method of
Ref[\onlinecite{maimaiti2019universal}] and showed how to generate
flat-band Hamiltonians with tunable winding number for the flat band,
and tunable gap and CLS configuration. We then used our construction
to study, in three-band systems, the effects on pairing and
superconductivity of the winding number, the CLS configuration, and
the gap between the flat band and the first band above it, for both
topolgical and non-topological flat bands. To this end we used both
full multi-band MF and exact DMRG calculations and found excellent
agreement between them especially in the topological cases. Our
results lead us to emphasize again that, in order to get accurate MF
results, it is crucial to consider both the order parameters and the
site densities as separate variational MF parameters on all
sublattices.

Specifically, we found for the gapped topological case, $W\neq 0$,
that, for low $U$ values, $D_s$ grows linearly with a slope which
depends sensitively on the choice of the CLS fillings even when $W$
and the integral over the BZ of the quantum metric remain fixed. The
optimal CLS choice (the one which gives the largest slope) is for
symmetric site densities on the two unit cells. Interestingly, if the
CLS is chosen with uniform site densities on all sites, the uniformity
breaks down for any $U\neq 0$: Sublattices $A$ and $C$ continue to
have the same density but not sublattice $B$. In the case of gapped
non-topological flat band, we find that $D_s$ is suppressed for $U$
smaller than the gap; it grows slowly as a power larger than $2$ until
$U$ is of the order of the band gap at which point band mixing helps
superconductivity.

In the gapless case, when the bands touch, we showed clear evidence of
a power law dependence $D_s \sim U^\varphi$ ($\varphi<1$), not the
logarithimic form $D_s \sim U {\rm ln}({\rm const.}/U)$, for both
topological and non-topological flat bands. Superconductivity at low
$U$ is therefore enhanced when the topological or non-topological flat
band touches the band above it.

In all cases we studied, the order parameter $\Delta^\alpha/U$
acquires large values for any finite $U$, as long as
$\rho^\alpha\neq0$. The phase difference of the order parameters was
found to depend on $\Phi$, $U$, and the band structure, unlike
previous two-band cases studied.\cite{chan2022pairing} Furthermore, we
showed that, as $U\to 0$, the single-particle correlation length
(extracted form the single particle Green's function) which is a
measure of the pair size, diverges as a power when the bands touch,
but tends to a constant less than one lattice spacing in the gapped
case. The pairs remain very small when the system is gapped, and their
size diverges in the gapless case --- but the divergence is slower
than in the dispersive band case where it is exponential.

Our results here offer insight into enhancement of superconductivity
in quasi-one dimensional systems and a methodology to finding
optimized Hamiltonians. With the increasing experimental ability to
realize designer systems governed by model Hamiltonians exhibiting a
flat band,\cite{xia16,zong16,mukherjee15,mukherjee17,baboux16} our
results can provide a practical road map.

The Hamiltonians we studied here do not have a chiral symmetry. It is
possible to construct such models\cite{maimaiti2019universal}, but due
to the chiral symmetry, the band structure must be symmetric about
zero energy. This means that if we want a flat band in the ground
state, the system must have an even number of orbitals. An example of
the two-band system (the Creutz lattice) was studied in
Ref.[\onlinecite{chan2022pairing}]. The next step would be to study
the four-band model (four coupled chains). Our multi-band MF method
can be easily applied in this situation, but exact DMRG calculations
will be rather challenging due to the large number of sites.

\underbar{\bf Acknowledgments:} S.M.C. is supported by a National
University of Singapore President's Graduate Fellowship. The DMRG
computations were performed with the resources of the National
Supercomputing Centre, Singapore (www.nscc.sg). This research is
supported by the National Research Foundation, Prime Minister's Office
and the Ministry of Education (Singapore) under the Research Centres
of Excellence programme. This project has also received funding from
Excellence Initiative of Aix-Marseille University - A*MIDEX, a French
“Investissements d’Avenir” program through the IPhU (AMX-19-IET-008)
and AMUtech (AMX-19-IET-01X) institutes.

\appendix

\section{Multi-band Mean Field Hamiltonian}\label{appendix:mf}

The full multi-band mean field is derived by decomposing the quartic
interaction term in the Hamiltonian with mean field parameters ---
$\Delta^\alpha$ and $\rho^\alpha_\sigma$ --- for each sublattice. As
in Ref.[\onlinecite{chan2022pairing}], but with three independent
sublattices, we write a trial Hamiltonian
\begin{equation}
\begin{aligned}
    H_{trial}&=H_K \\
    &-U\displaystyle\sum_{j,\alpha}\rho^\alpha_\uparrow
    c^{\alpha\dagger}_{j,\downarrow}c^{\alpha}_{j,\downarrow}+\rho^\alpha_\downarrow
    c^{\alpha\dagger}_{j,\uparrow}c^{\alpha}_{j,\uparrow}\\ 
    &-\displaystyle\sum_{j,\alpha}\Delta^\alpha
    c^{\alpha\dagger}_{j,\downarrow}c^{\alpha\dagger}_{j,\uparrow} +
    \Delta^{\alpha*}c^{\alpha}_{j,\uparrow}c^{\alpha}_{j,\downarrow},\\ 
    H_K&=\displaystyle\sum_{i,j,\alpha,\sigma}\left(
    t_{ij}^{\alpha,\alpha'}c^{\alpha\dagger}_{i,\sigma}
    c^{\alpha'}_{j,\sigma} +h.c.\right) \\
    &-\mu\displaystyle\sum_{j,\alpha,\sigma}
    c^{\alpha\dagger}_{j,\sigma} c^{\alpha}_{j,\sigma},
\end{aligned}
\end{equation}
with mean field parameters $\Delta^\alpha$ and
$\rho^\alpha_\sigma$. The Gibbs-Bogoliubov inequality
\cite{kuzemsky2015variational} gives,
\begin{equation}
\begin{aligned}
    F&\leq F_{trial}  \\
    &-\left\langle  U\displaystyle\sum_{j,\alpha}
    c^{\alpha\dagger}_{j,\downarrow} c^{\alpha\dagger}_{j,\uparrow}
    c^{\alpha}_{j,\uparrow}c^{\alpha}_{j,\downarrow}\right\rangle_{trial}
    \\ 
    &+\left\langle U\displaystyle\sum_{j,\alpha}\rho^\alpha_\uparrow
    c^{\alpha\dagger}_{j,\downarrow}c^{\alpha}_{j,\downarrow}+\rho^\alpha_\downarrow
    c^{\alpha\dagger}_{j,\uparrow}c^{\alpha}_{j,\uparrow}\right\rangle_{trial}\\
    &+\left\langle\displaystyle\sum_{j,\alpha}\Delta^\alpha
    c^{\alpha\dagger}_{j,\downarrow}c^{\alpha\dagger}_{j,\uparrow}+
    \Delta^{\alpha*}c^{\alpha}_{j,\uparrow}c^{\alpha}_{j,\downarrow}
    \right\rangle_{trial},
    \end{aligned}
\end{equation}
where $\langle \dots \rangle_{trial}$ denotes expectation values with
respect to the weight ${\rm e}^{-\beta H_{trial}}/Z_{trial}$ with
$Z_{trial}={\rm Tr}\,{\rm e}^{-\beta H_{trial}}={\rm e}^{-\beta
  F_{trial}}$. Minimizing the right hand side with respect to the MF
variational parameters, we obtain an upper bound on the true free
energy, which we define as the mean field free energy, $F_{MF}$:
\begin{equation}
    F_{MF}=F_{trial}+UL\displaystyle\sum_\alpha \left(
    \rho^\alpha_\uparrow \rho^\alpha_\downarrow +
    \abs{\Delta^\alpha/U}^2 \right),
\end{equation}
where $F_{MF}=\langle H_{MF} \rangle$ and $F_{trial}=\langle H_{trial}
\rangle$ at $T=0$. The mean field parameters can be expressed,
following the optimization, as
\begin{equation}
    \begin{aligned}
    \rho^\alpha_\sigma&=\langle c^{\alpha\dagger}_{j,\sigma}
    c^{\alpha}_{j,\sigma}\rangle, \\
    \Delta^\alpha&=U\langle
    c^{\alpha}_{j,\uparrow}c^{\alpha}_{j,\downarrow} \rangle.
    \end{aligned}
\end{equation}
This defines $H_{MF}$ to be
\begin{equation}
    \begin{aligned}
    H_{MF}&= \displaystyle\sum_{i,j,\alpha,\sigma}\left(
    t_{ij}^{\alpha,\alpha'}c^{\alpha\dagger}_{i,\sigma}c^{\alpha'}_{j,\sigma}
    + h.c.\right)\\
    &-U\displaystyle\sum_{j,\alpha}\rho^\alpha_\uparrow
    c^{\alpha\dagger}_{j,\downarrow}c^{\alpha}_{j,\downarrow} +
    \rho^\alpha_\downarrow c^{\alpha\dagger}_{j,\uparrow}
    c^{\alpha}_{j,\uparrow}\\
    &-\displaystyle\sum_{j,\alpha}\Delta^\alpha
    c^{\alpha\dagger}_{j,\downarrow}c^{\alpha\dagger}_{j,\uparrow} +
    \Delta^{\alpha*}c^{\alpha}_{j,\uparrow}c^{\alpha}_{j,\downarrow}\\
    &-\mu\displaystyle\sum_{j,\alpha,\sigma}
    c^{\alpha\dagger}_{j,\sigma}c^{\alpha}_{j,\sigma}\\
    &+L\displaystyle\sum_\alpha
    U\rho^\alpha_\uparrow\rho^\alpha_\downarrow +
    \abs{\Delta^\alpha}^2/U.
    \end{aligned}
\end{equation}
With equal population of $\uparrow$- and $\downarrow$-spins, we can
choose to replace
$\rho^\alpha_\uparrow=\rho^\alpha_\downarrow=\rho^\alpha$. To study
the superfluid behavior of the system, we apply a phase twist $\Phi$
with $c^{\alpha}_{j,\sigma}\rightarrow
c^{\alpha}_{j,\sigma}e^{ij\Phi/L}$. In general, we can write the
Fourier transformed mean-field Hamiltonian with a phase gradient as
\begin{equation}
\begin{aligned}
H_{MF}(\Phi) &= \displaystyle\sum_k \Psi^\dagger_k \mathcal{M}_k
(\Phi)\Psi_k \\
&+L\displaystyle\sum_\alpha\left( U\rho^\alpha_\uparrow
\rho^\alpha_\downarrow +
\abs{\Delta^\alpha}^2/U-U\rho^\alpha_\downarrow -\mu\right)
\end{aligned}    
\end{equation}
with $\Psi_k^\dagger=\begin{pmatrix} c_{k\uparrow}^{A\dagger}
&c_{k\uparrow}^{B\dagger} &c_{k\uparrow}^{C\dagger}
&c_{-k\downarrow}^A &c_{-k\downarrow}^B
&c_{-k\downarrow}^C \end{pmatrix}$ the Nambu spinor, and the block
matrix
\begin{equation}
    \mathcal{M}_k(\Phi)=\begin{pmatrix}
    \mathcal{K}(\phi+k) & \mathcal{D} \\
    \mathcal{D}* & -\mathcal{K}^T(\phi-k)
    \end{pmatrix}.
\end{equation}
The block $\mathcal{D}$ simply takes into account the pairing order
parameter,
\begin{equation}
    \mathcal{D} = 
\begin{pmatrix}
\Delta^A & 0 & 0 \\
0 & \Delta^B & 0 \\
0 & 0 & \Delta^C
\end{pmatrix}.
\end{equation}
The block $\mathcal{K}(\phi\pm k)$ expresses the hopping terms and
sublattice-dependent filling as a modification to the chemical
potential 
\begin{equation}
    \mathcal{K}(\phi+k)=\begin{pmatrix}
    K_{11}-\Tilde{\mu}^A & K_{12} & K_{13} \\
     K_{21} & K_{22}-\Tilde{\mu}^B & K_{23} \\
     K_{31} & K_{32} & K_{33}-\Tilde{\mu}^C
    \end{pmatrix},
\end{equation}
where the sublattice-dependent chemical potential
$\Tilde{\mu}^\alpha=\mu+\rho^\alpha U$ is crucial to describe
accurately a system with non-identical
sublattices. $K_{11}=2t_a\cos(\phi+k)$, $K_{22}=2t_e\cos(\phi+k)$,
$K_{33}=2t_i\cos(\phi+k)$ are inter-cell hopping terms on the same
sublattice. $K_{12}=t_1+t_b e^{i(\phi+k)}+t_d
e^{-i(\phi+k)}=K_{21}^*$, $K_{23}=t_2+t_f e^{i(\phi+k)}+t_h
e^{-i(\phi+k)}=K_{32}^*$, $K_{13}=t_3+t_c e^{i(\phi+k)}+t_g
e^{-i(\phi+k)}=K_{31}^*$.

\section{Construction of Flat Band Hamiltonians}

\label{appendix:fbh}
We now outline the method to construct Hamiltonians with a flat band
in the ground state\cite{maimaiti2019universal} and show how to fix
the windining number. The most general form of the CLS on two unit
cells is
\begin{equation}
\label{CLS_general}
    \Psi_1 = 
    \begin{pmatrix} a \\
    be^{i\beta}\\
    ce^{i\gamma}
    \end{pmatrix}, \Psi_2 = 
    \begin{pmatrix} xe^{i\chi} \\
    ye^{i\tau}\\
    ze^{i\zeta}
    \end{pmatrix}
\end{equation}

The first condition is to have\cite{maimaiti2019universal}

\begin{equation}
    \bra{\Psi_1}\ket{\Psi_2}= axe^{i(\chi)} + bye^{i(\tau-\beta)} +
    cze^{i(\zeta-\gamma)}=1.
\end{equation}
We obtain the Bloch state by Fourier transforming, which can be
written as
\begin{equation}
    \ket{\Psi_k}=\frac{1}{R}\begin{pmatrix}
    a+xe^{i(\chi-k)} \\
    be^{i\beta} + ye^{i(\tau-k)} \\
    ce^{i\gamma} + ze^{i(\zeta-k)}
    \end{pmatrix}.
\end{equation}
To normalize the Bloch state, the expression which gives R is
\begin{equation}
    R^2 = 2\cos(k)+a^2+b^2+c^2+x^2+y^2+z^2.
\end{equation}
Differentiating the normalized Bloch state with respect to the lattice
momentum, we obtain

\begin{equation}
    \begin{aligned}
    \ket{\partial_k\Psi_k} &= \frac{1}{R^3}\begin{pmatrix}
    (a+xe^{i(\chi-k)})\sin(k)-iR^2xe^{i(\chi-k)} \\
    (be^{i\beta}+ye^{i(\tau-k)})\sin(k)-iR^2ye^{i(\tau-k)} \\
    (ce^{i\gamma}+ze^{i(\zeta-k)})\sin(k)-iR^2ze^{i(\zeta-k)}
    \end{pmatrix}.
    \end{aligned}
\end{equation}
The winding number is given by,
\begin{equation}
    \begin{aligned}
    &W\pi=i\int_0^{2\pi}dk\bra{\Psi_k}\ket{\partial_k\Psi_k}, \\
    &=\frac{1}{2}\int_0^{2\pi}dk(1+\frac{x^2+y^2+z^2-a^2-b^2-c^2}{2\cos(k)
      +a^2+b^2+c^2+x^2+y^2+z^2}), \\
\label{eq:Wtune}
    &=\pi+\frac{\pi(x^2+y^2+z^2-a^2-b^2-c^2)}{\sqrt{(a^2+b^2+c^2 +
        x^2+y^2+z^2)^2-4}}.
    \end{aligned}
\end{equation}
We see that by choosing $\Psi_1$ and $\Psi_2$ appropriately, one can
tune to the desired value of $W$. The integral over the Brillouin zone
of the quantum metric can now be expressed as \widetext
\begin{equation}
        \mathcal{Q}=\frac{1}{2\pi}\displaystyle\int_0^{2\pi}\Re(g(k))dk
        =\frac{[a^2+b^2+c^2+x^2+y^2+z^2][ (a^2+b^2+c^2)^2 +
            (x^2+y^2+z^2)^2 -2]}{\left((a^2+b^2+c^2+x^2+y^2+z^2)^2 -
          4\right)^\frac{3}{2}}.
\end{equation}
\twocolumngrid With $g(k)= 2(\bra{\partial_k\Psi_k}
\ket{\partial_k\Psi_k}- \abs{(\bra{\Psi_k} \ket{\partial_k\Psi_k}}^2)$
the quantum geometric tensor, and its real part the quantum metric. To
find the hopping potentials, we choose a flat band energy $E_{FB}$ and
satisfy the conditions in Ref.[\onlinecite{maimaiti2019universal}].

For real CLS, the two equations to solve for intra-cell hopping terms
are,
\begin{equation}
\label{EFB}
\begin{aligned}
    E_{FB}&=\bra{\Psi_2}H_0\ket{\Psi_1},\\
    &=t_1(bx+ay)+t_2(cy+bz)+t_3(cx+az),
\end{aligned}
\end{equation}
and
\begin{equation}
    \begin{aligned}
    \bra{\Psi_1}E_{FB}-H_0\ket{\Psi_1}&= \bra{\Psi_2} E_{FB}-H_0
    \ket{\Psi_2}.
    \end{aligned}
\end{equation}

The expression for $H_1$ is 
\begin{equation}
\label{H1_expression}
    H_1 = \frac{(E_{FB}-H_0)\ket{\Psi_1} \bra{\Psi_2} (E_{FB}-H_0)}
    {\bra{\Psi_1} E_{FB} - H_0 \ket{\Psi_1}}+Q_{12}KQ_{12}.
\end{equation}
The term $K$ is arbitrary and $Q_{12}$ is constructed from the CLS.
\begin{equation}
    \begin{aligned}
    Q_{12}&=R_{12}Q_1, \\
    Q_{i} &=
    \mathbb{I}-\frac{\ket{\Psi_{i}}\bra{\Psi_{i}}}{\bra{\Psi_{i}}
      \ket{\Psi_{i}}}, \\
    R_{12} &= \mathbb{I}-\frac{Q_1\ket{\Psi_{2}} \bra{\Psi_{2}}Q_1}
    {\bra{\Psi_{2}}Q_1\ket{\Psi_{2}}}.
    \end{aligned}
\end{equation}

In general, for a chosen winding number, there is thus an infinite
number of flat band lattices that can be constructed.

Equation (\ref{eq:Wtune}) shows how to tune the winding number to a
desired value. To obtain $W=0$, we obtain the condition
\begin{equation}
    \abs{\Psi_1}^2\abs{\Psi_2}^2=(a^2+b^2+c^2)(x^2+y^2+z^2)=1.
\end{equation}
This gives the normalization condition
\begin{equation}
    \bra{\Psi_1}\ket{\Psi_2} =1=\abs{\Psi_1}\abs{\Psi_2}\cos(\theta)
    =\cos(\theta),
\end{equation}
which implies that $\ket{\Psi_1}= M\ket{\Psi_2}$, where $M$ is a
constant. M can simply be expressed through
\begin{equation}
    \ket{\Psi_1}=\ket{\Psi_2}\bra{\Psi_1}\ket{\Psi_1}.
\end{equation}
From Eq.(\ref{H1_expression}), we consider the denominator
$\bra{\Psi_1}E_{FB}-H_0\ket{\Psi_1}$,
\begin{equation}
    \begin{aligned}
    &\bra{\Psi_1}E_{FB}-H_0\ket{\Psi_1}, \\
    &= E_{FB}\bra{\Psi_1}\ket{\Psi_1}-\bra{\Psi_1}H_0\ket{\Psi_1},\\
    &=E_{FB}\bra{\Psi_1}\ket{\Psi_1}-\bra{\Psi_1}H_0\ket{\Psi_2}
      \bra{\Psi_1}\ket{\Psi_1}, \\
    &=E_{FB}\bra{\Psi_1}\ket{\Psi_1}-E_{FB}\bra{\Psi_1}\ket{\Psi_1}, \\
    &=0.
    \end{aligned}
\end{equation}
This means that for a finite $H_1$, we have an additional condition
that $(E_{FB}-H_0)\ket{\Psi_1}\bra{\Psi_2}(E_{FB}-H_0)=0$. To
determine $H_1=Q_{12}KQ_{12}$, we can work out that
$Q_1\ket{\Psi_2}=0$ and $R_{12}=\mathbb{I}$. To show that the CLS is
localized within one unit cell,
\begin{equation}
    H_1\ket{\Psi_1} = Q_1KQ_1\ket{\Psi_1} = 0.
\end{equation}
For a lattice with $W=0$ and a chosen real CLS and $E_{FB}$, there is
a unique solution for the intra-cell hopping terms, and we can only
have the CLS localized within one unit cell.

\section{Band Structure and Order Parameters}
\label{appendix:Bands_Deltas}

\subsection{Lattice $\mathcal{A}$}

\begin{figure}[h!]
 \includegraphics[width=8.5cm]{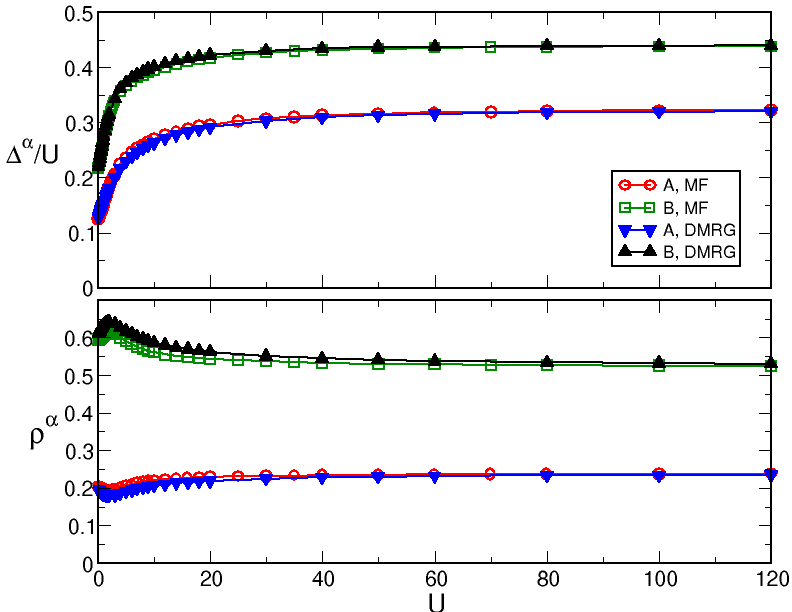}
 \caption{(Color online) The order parameter and sublattice fillings
   for the isolated flat band of Lattice $\mathcal{A}$ ($\kappa=0$)
   with $\rho=1/3$. Top: Order parameters acquire a large nonzero
   value for any finite $U$, and $\Delta^A/U=\Delta^C/U$ for
   $\Phi=0$. Bottom: $\rho^\alpha$ are shown to be sublattice
   dependent, and this can only be properly reproduced with MF when
   $\rho^\alpha$ are considered as MF parameters.}
 \label{fig:gapCLS-A_DeltaDens}
\end{figure}

\begin{figure}[h!]
 \includegraphics[width=8cm]{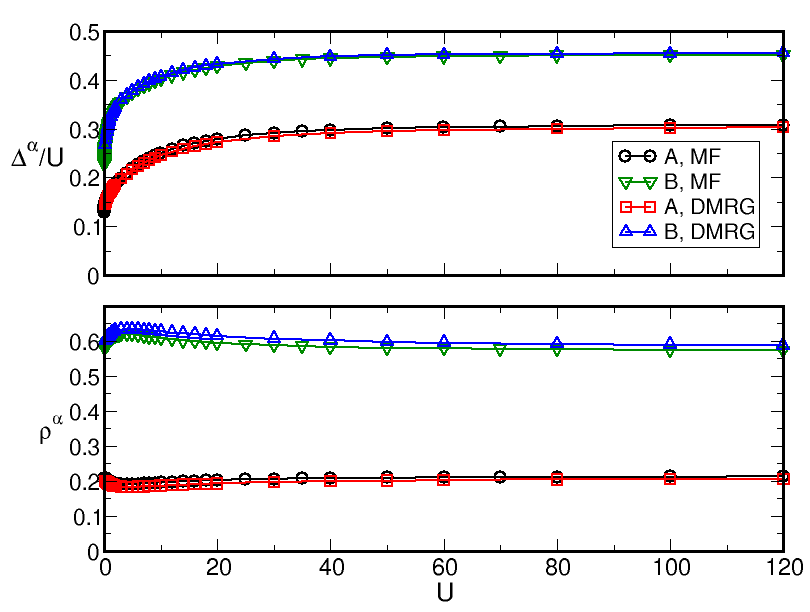}
 \caption{(Color online) The order parameter and sublattice fillings
   for the non-isolated flat band of Lattice $\mathcal{A}$
   ($\kappa=-0.375$) with $\rho=1/3$. Top: Order parameters
   are sublattice depedendent with $\Delta^A/U=\Delta^C/U$ for
   $\Phi=0$. Bottom: $\rho^\alpha$, sublattice
   fillings.}
 \label{fig:gaplessCLS-A_DeltaDens}
\end{figure}

Figures \ref{fig:gapCLS-A_DeltaDens} and
\ref{fig:gaplessCLS-A_DeltaDens} show the agreement between DMRG and
full MF for the order parameters and sublattice fillings for both the
gapped and gapless cases of Lattice $\mathcal{A}$ ($W=1$) studied in
Section \ref{results:nonzero_winding}. Note that the order parameters
$\Delta^\alpha/U$ acquire a large finite value once $U$ is nonzero,
while $\Delta^\alpha\propto U$ linearly. An important point to note is
the sublattice fillings, where sublattices A and C have equal filling
but sublattice B is different. This is only faithfully reproduced in
the MF that we propose, and not when BCS MF is employed (Appendix
\ref{appendix:MFwithoutrhos}).
\subsection{Lattice $\cal B$}
\begin{figure}
    \includegraphics[width=8.6cm]{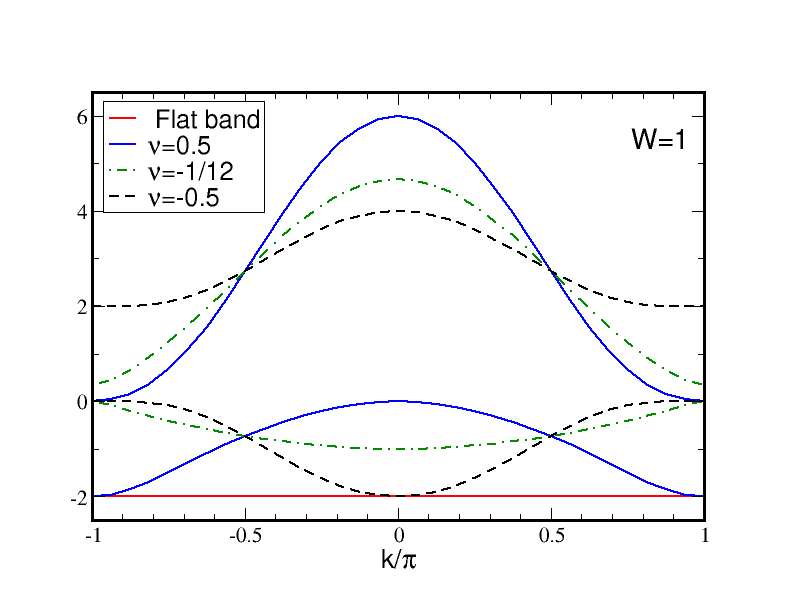}
    \caption{Band structure of Lattice $\mathcal{B}$}
    \label{fig:Bandstructure-CLS-B}
\end{figure}
The band structures of Lattice $\mathcal{B}$ ($W=1$), for cases where
we computed $D_s$ (Figure \ref{fig:DsvsU-CLS-B}) are shown in
Fig. \ref{fig:Bandstructure-CLS-B}. The bands touch at $k=0$ for
$\nu=-0.5$ and $k=\pi$ for $\nu=0.5$; the gapped case with $E_{gap}=1$
has $\nu =-1/12$.

\subsection{Lattice $\mathcal{C}$}
\begin{figure}[h]
    \includegraphics[width=8.5cm]{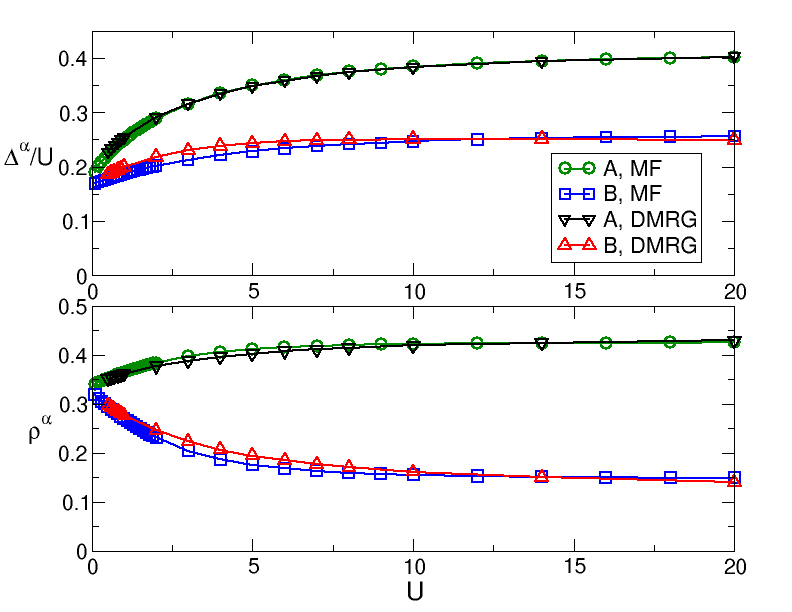}
    \caption{Pairing order parameter and sublattice fillings for
      Lattice $\mathcal{C}$ ($W=0$), touching bands with
      $\varrho=1.5$. While the values of $D_s$ calculated using MF and
      DMRG methods agree well qualitatively and less well
      quantiatively, the order parameter and sublattice fillings are
      accounted for very well by the full mean field method. ($\rho =
      1/3$)}
    \label{fig:DeltaDens-CLS-C}
\end{figure}
In Fig. \ref{fig:DeltaDens-CLS-C}, we show, for lattice ${\cal C}$
touching bands with $\varrho=1.5$, that although $D_s$ computed with
MF and DMRG agree well qualitatively but less so quantitatively for
the non-topological flat bands ($W=0$), the pairing order parameter
and sublattice fillings are modelled very well by the full mean
field. The agreement between MF and DMRG for $\Delta^\alpha$ and
$\rho^\alpha$ was also observed for the isolated non-topological flat
band.

\section{Other lattices}\label{appendix:OtherLattices}
\begin{figure}[h!]
 \includegraphics[width=8.6cm]{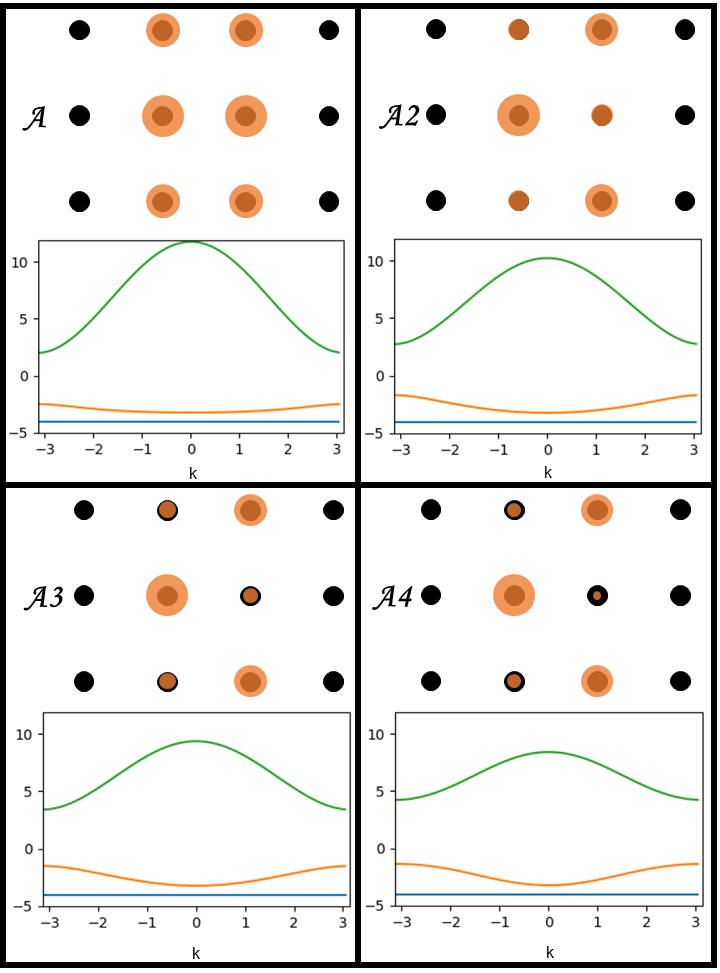}
 \caption{(Color online) CLS is increasingly asymmetrical from lattice
   $\cal A$ to $\mathcal{A}4$. The flat band energy is kept constant
   at $-4$ and the band gap is $E_{gap}\approx0.7625$, and $W=1$. }
 \label{fig:varyCLS}
\end{figure}
\begin{figure}[h!]
 \includegraphics[width=8.6cm]{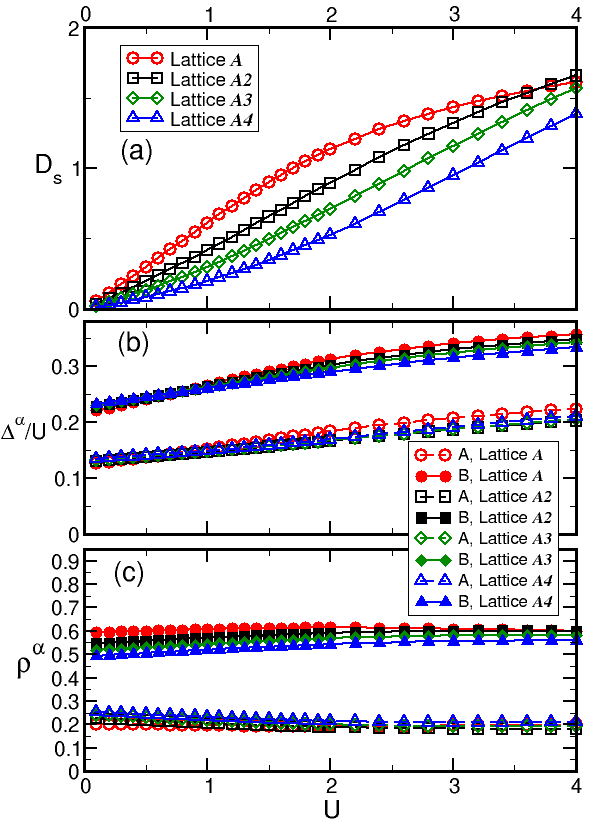}
 \caption{(Color online) $D_s$, $\rho^\alpha$ and $\Delta^\alpha$ for
   lattices $\mathcal{A}, \mathcal{A}2, \mathcal{A}3, \mathcal{A}4$
   (Fig. \ref{fig:varyCLS}) obtained through MF computation. The
   pairing order parameters, site fillings and integral over the
   BZ of the quantum metric vary slightly, while the superfluid
   density is evidently distinct for the cases considered. ($\rho =
   1/3$)}
 \label{fig:DsvsU-CLSvarious}
\end{figure}
Here, we show that $D_s$ is strongly dependent on the CLS.  We
consider lattice $\cal A$ ($W=1$) and tune the CLS while keeping
$E_{FB}=-4, E_{gap}\approx 0.7625, t_2=\sqrt{7}$ and $t_3=1$ constant,
through the change of $H_1$ and $t_1$. The CLS and band structures
that we consider are shown in Fig. \ref{fig:varyCLS}, labeled lattices
$\mathcal{A}, \mathcal{A}2, \mathcal{A}3, \mathcal{A}4$, with
increasing asymmetry of the CLS.

We compute $D_s$, sublattice fillings and order parameters with MF for
these lattices, and show that while $\rho^\alpha$ and $\Delta^\alpha$
do not vary significantly (Fig. \ref{fig:DsvsU-CLSvarious} (b) and
(c)), $D_s$ changes substantially across the cases considered,
Fig. \ref{fig:DsvsU-CLSvarious} (a). In general, the slope is largest
for the most symmetric CLS, which also has the fastest exponential
decay of the Wannier function (not shown). This hints towards
optimizing superconductivity by engineering the Hamiltonian which
corresponds to the most symmetric localized state. We emphasize that
while the winding number is constant, and the integral over the
BZ of the quantum metric does not vary much ($\mathcal{Q_A}=0.505$,
$\mathcal{Q_A}_2=0.507$, $\mathcal{Q_A}_3=0.509$,
$\mathcal{Q_A}_4=0.516$), the slopes of $D_s$,
Fig. \ref{fig:DsvsU-CLSvarious}(a), are evidently
distinct. Additionally, for lattice $\cal B$ with symmetric CLS and
equal filling on all CLS sites, the slope at low $U$ is equal to that
of lattice $\cal A$.
\begin{figure}[h!]
 \includegraphics[width=8.6cm]{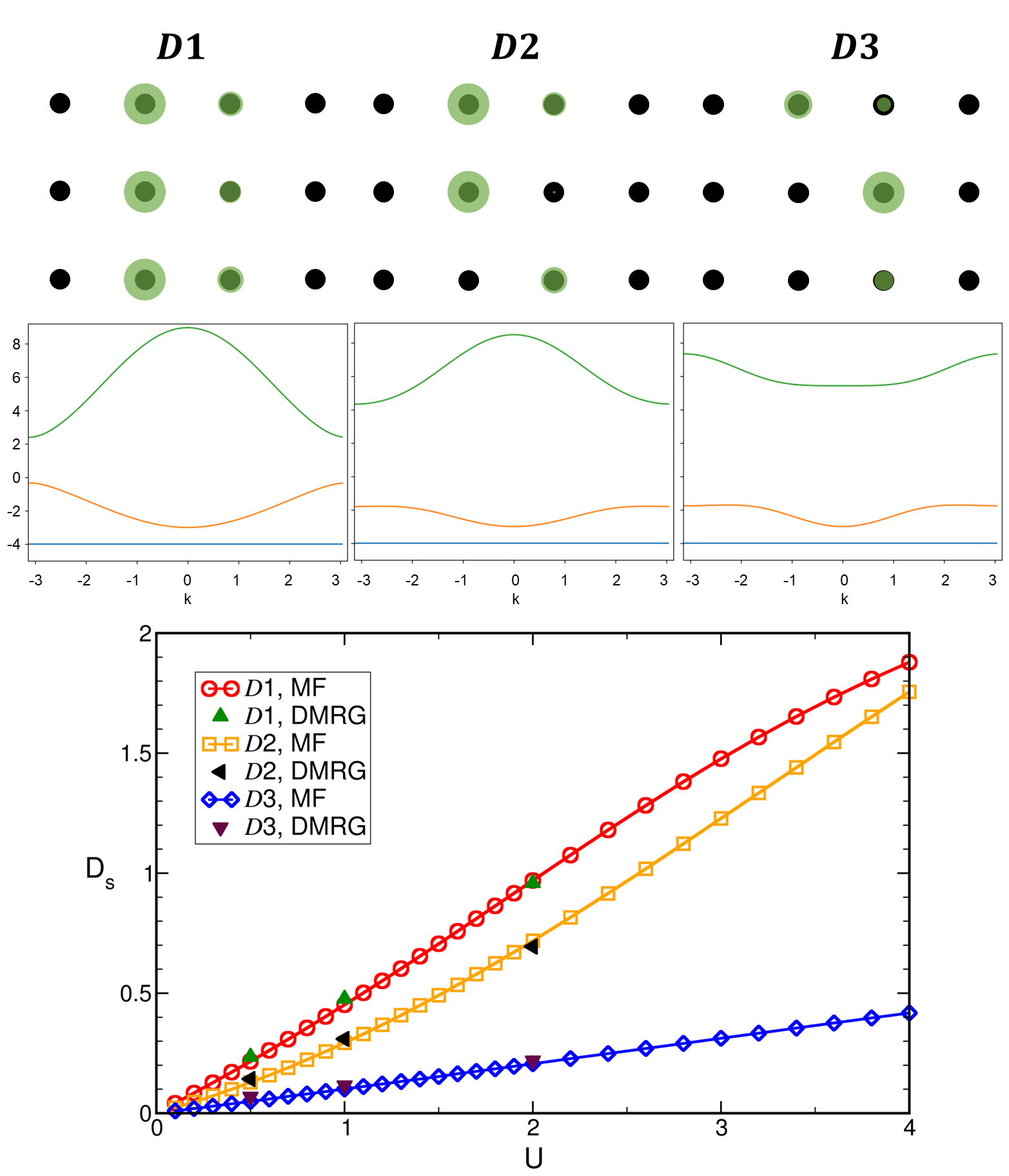}
 \caption{(Color online) $D_s$ vs $U$ for $W=0.5$, $\rho=1/3$, with
   corresponding band structures and CLS depicting various
   asymmetries. We include DMRG points showing the agreement of the
   full MF at finite, non-integer winding.}
 \label{fig:DsvsUw0.5}
\end{figure}

The MF and DMRG agreement extends to lattices with non-integer, finite
winding. Again, with equal $E_{gap}=1$, $E_{FB}=-4$ and similar band
structures, we show that $D_s$ is dependent on the CLS. For a winding
of $W=0.5$, we obtain asymmetric CLS and give three examples: lattices
$\mathcal{D}1, \mathcal{D}2$ and $\mathcal{D}3$. Note that for
$W=0.5$, the CLS cannot be symmetric but the degree of the asymmetry
can be tuned. To eliminate the effects contributed through the
uppermost band, we focus on the range $0<U\leq 4$. We propose that the
optimization of $D_s$ is contingent on occupation of CLS, where one
should identify the model with filling most symmetric on all
sublattices. This means that if some sublattice occupations of the CLS
are zero, $D_s$ will be significantly reduced.

\section{Failing of MF without $\rho^\alpha$ as a MF parameter}\label{appendix:MFwithoutrhos}

We have insisted repeatedly on the importance of including the
site-dependent fillings as MF parameters. Here we show how the BCS
approach fails when taken with the correct complex order parameters
but without the site densities. In the BCS MF approach, which has been
extensively employed by many
studies\cite{julku2016geometric,huhtinen2022revisiting,iskin2019origin,wu2021superfluid},
the Hubbard interaction term is simply decomposed as

\begin{equation}
\begin{aligned}
-U c^{\alpha\dagger}_{j,\downarrow} c^{\alpha\dagger}_{j,\uparrow}
c^{\alpha}_{j,\uparrow} c^{\alpha}_{j,\downarrow} = -\Delta^\alpha
c^{\alpha\dagger}_{j,\downarrow} c^{\alpha\dagger}_{j,\uparrow} +
\Delta^{\alpha*} c^{\alpha}_{j,\uparrow} c^{\alpha}_{j,\downarrow} +
\frac{\abs{\Delta^\alpha}^2}{U}.
\end{aligned}
\end{equation}

In general, for the BCS MF, the site densities will go to the same
value $\rho^\alpha\rightarrow\rho$ as $U$ increases. As a result, the
order parameters $\Delta^\alpha$ for the sublattices will also tend to
the same value at large $U$. This is both qualitatively and
quantitatively wrong, as we have presented in Appendix
\ref{appendix:Bands_Deltas}, where sublattices continue to be
inequivalent even at large $U$.

\begin{figure}[h!]
 \includegraphics[width=8cm]{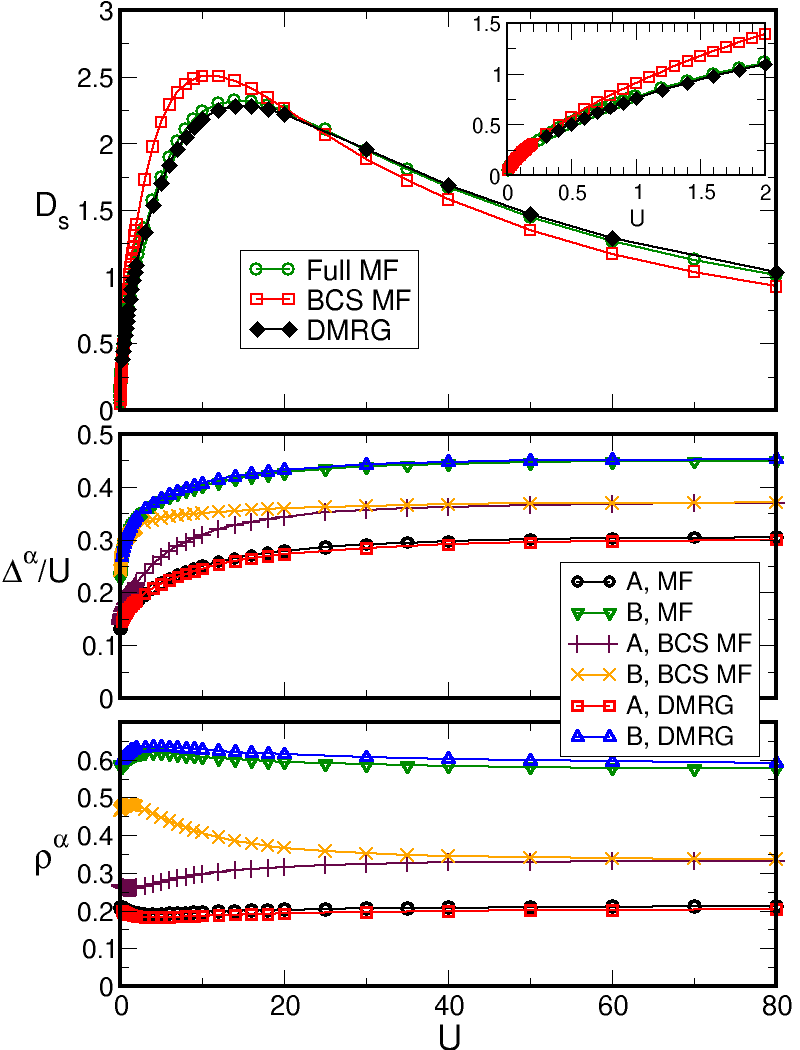}
 \caption{(Color online) Lattice $\mathcal{A}$ ($W=1$) gapless at
   $k=0$ ($\kappa=-0.375$): BCS MF approach without sublattice
   densities as mean field parameters compared with the full MF and
   DMRG shows a clear disagreement. Top: $D_s$ as a function of $U$
   obtained from BCS MF agrees for very weak interactions (inset) but
   quickly deviates from full MF and DMRG results. The discrepancy is
   especially prominent in the middle panel: the order parameter and
   the bottom panel: the site fillings, where BCS MF inaccurately
   models sublattices with equivalent $\Delta^\alpha$ and
   $\rho^\alpha$ at large $U$. The filling is $\rho = 1/3$.}
 \label{fig:gaplessCLS-A_BCS}
\end{figure}

As an example, we show in Fig. \ref{fig:gaplessCLS-A_BCS} a comparison
of full multi band MF, BCS MF and DMRG results for the gapless case
($\kappa=-0.375$) of Lattice $\cal A$ ($W=1$). At weak coupling, we
find that the BCS MF does not capture the actual behavior of $D_s$
well, with the inaccurate power dependence evident in the inset. As
$U$ increases, the disagreement becomes increasingly apparent.
\newline

Consequently, while one can focus on the superfluid weight (and its
relation to the quantum metric) calculated through the BCS MF and
argue that it is an acceptable agreement, a closer look at the
sublattice equivalence and properties reveals the break down of this
approach.  \bibliography{3bandrefs.bib}

\end{document}